\documentclass[aps,preprint,a4paper,nofootinbib,superscriptaddress]{revtex4-1}
\usepackage{dsfont}
 \usepackage{graphicx}
\usepackage{longtable}
\usepackage{amsmath,amssymb}

\newcommand{\de}{\mathrm{d}}
\newcommand{\De}{\mathcal{D}}

\newcommand{\pa}{\partial}
\newcommand{\qq}{\,,\quad}
\newcommand{\R}{\mathds{R}}
\newcommand{\N}{\mathcal{N}}

\newcommand{\ep}{\epsilon}
\newcommand{\tr}{\textrm{Tr}\,}

\newcommand{\wt}[1]{\widetilde{#1}}
\newcommand{\pexp}{\mathcal{P }\exp}
\newcommand{\reff}[1]{(\ref{#1})}

\newcommand{\be}{\begin{equation}}
\newcommand{\ee}{\end{equation}}
\newcommand{\nn}{\nonumber}

\begin{document} 
\title{Correlators of Hopf Wilson loops in the AdS/CFT correspondence}
\author{Luca Griguolo}
\affiliation{Dipartimento di Fisica, Universit\`a di Parma and INFN Gruppo Collegato di Parma, \\ Viale G.P. Usberti 7/A, 43100 Parma, Italy}  
\author{Stefano Mori}
\affiliation{Dipartimento di Fisica, Universit\`a di Parma and INFN Gruppo Collegato di Parma, \\ Viale G.P. Usberti 7/A, 43100 Parma, Italy}  
\author{Fabrizio Nieri}
\affiliation{Dipartimento di Fisica e Astronomia, Universit\`a di
Firenze and INFN Sezione di Firenze,\\ Via  G. Sansone 1, 50019 Sesto Fiorentino, Italy} 
\author{Domenico Seminara}
\affiliation{Dipartimento di Fisica e Astronomia, Universit\`a di
Firenze and INFN Sezione di Firenze,\\ Via  G. Sansone 1, 50019 Sesto Fiorentino, Italy} 

\email{luca.griguolo@fis.unipr.it} 
\email{stefano.mori@fis.unipr.it} 
\email{nieri@fi.infn.it} 
\email{seminara@fi.infn.it}
\begin{abstract}We study at quantum level correlators of supersymmetric Wilson loops with contours lying on Hopf fibers of $S^3$. In $\mathcal{N}=4$ SYM theory the strong coupling analysis can be performed using the AdS/CFT correspondence and a connected classical string surface, linking two different fibers, is presented. More precisely, the string solution describes oppositely oriented fibers with the same scalar coupling and depends on an angular parameter, interpolating between a non-BPS configuration and a BPS one. The system can be thought as an alternative deformation of the ordinary antiparallel lines giving the static quark-antiquark potential, that is indeed correctly reproduced, at weak and strong coupling, as the fibers approach one another.
\end{abstract}
\maketitle

\maketitle

\section{Introduction}
The Wilson loop is one of the most important observable in gauge theories and it has been suspected for a long time that its dynamics should have a natural description in terms of strings. The AdS/CFT correspondence provides a concrete realization of this idea: the quantum expectation value of Wilson loops in $\N=4$ Super Yang-Mills theory (SYM$_{4}$) can be computed by means of the partition function of the type IIB string in $AdS_5\times S^5$, satisfying suitable boundary conditions \cite{pot1,mwloop,dgo}. Considering $U(N)$ as the gauge group and denoting the coupling constant by $g$, the large $N$ and strong-coupling $\lambda\equiv g^2N$ limit is equivalent to a semiclassical approximation and the computation reduces to find the minimal area surfaces in $AdS_5\times S^5$, with boundary conditions imposed by the Wilson loop operators. 

\noindent
In addition to the familiar interaction with the gauge field, Wilson loops in SYM$_{4}$ naturally contain scalar fields coupled to the contour, a presence that can be easily traced back from both sides of the correspondence \cite{pot1,mwloop,dgo}. It turns out that for suitable choices of the scalar couplings, as functions of the loop variables, the resulting Wilson loop appears to be $supersymmetric$, i.e. invariant with respect to some of the supersymmetry transformations of the theory. A general class of such operators was proposed by Zarembo \cite{Zarembo:2002an} and have trivial expectation value. A second family of supersymmetric Wilson loops (that we denote as DGRT), having instead a non-trivial dependence on the gauge coupling constant, was constructed in \cite{dgrta}, the well studied circular loop 
\cite{eszcircularloop,drukkergrossexact} being a particular representative of this ensemble. The contours lay, in general, into a sphere $S^3$ embedded into the Euclidean $\mathbb{R}^4$ and the operators are generically 1/16 BPS. The maximal supersymmetric configuration is the 1/2 BPS circular case, whose exact expectation value is captured by a Gaussian matrix model 
\cite{eszcircularloop,drukkergrossexact} as result of a localization procedure \cite{matrixmodelloc}. When restricted on $S^2$ the loops are at least 1/8 BPS and a tight relation with two-dimensional Yang-Mills theory on the sphere has been observed \cite{dgrta,Young:2008ed,Bassetto:2008yf,Giombi:2009ms,Bassetto:2009rt,Bassetto:2009ms,Giombi:2009ds,Pestun:2009nn}. Recently a systematic classification of supersymmetric Wilson loops invariant under at least one superconformal symmetry has been presented \cite{Dymarsky:2009si}: weak and strong coupling computations for some new loops appear in 
\cite{Cardinali:2012sy}.

Supersymmetric Wilson loops are important objects to study for many reasons. For example, they provide non-trivial tests on the validity of the AdS/CFT correspondence: due to their BPS nature, one can hope to obtain exact results in the gauge theory to be compared with the strong coupling predictions on the string side. For the 1/2 BPS circular loop the exact quantum field theoretical result is indeed reproduced by the classical string computation \cite{eszcircularloop,drukkergrossexact}, while the first subleading term was checked in \cite{Kruczenski:2008zk}. More recently the same limit has been performed, starting from the matrix model obtained by localization in  \cite{matrixmodelloc}, for the BPS circular loop in ${\cal N}=2$ Superconformal QCD, suggesting a logarithmic behavior of the effective string tension on the 't Hooft coupling \cite{Passerini:2011fe}. On the other hand, also non-supersymmetric configurations are of great importance. For instance, as is well known, the non-BPS antiparallel lines allow to define and compute the static quark-antiquark potential 
\cite{pot1,mwloop,pot2,Pineda:2007kz} and the strong coupling limit can be carefully studied at subleading level 
\cite{Chu:2009qt,Forini:2010ek}. More recently a generalized configuration \cite{Drukker:2011za}, interpolating between BPS and non-BPS operators, has been introduced and studied both at weak and strong coupling level: in this framework a new non-supersymmetric observable can be defined and exactly computed \cite{Correa:2012at,Correa:2012hh,Drukker:2012de}.

Configurations involving more than one circuit at once (usually two) define Wilson loop correlators. These objects are also very interesting since they carry information about loop interactions: on the string theory side of the correspondence there is the very suggestive picture that a connected correlator is described by a minimal area surface linking the boundary loops \cite{grossooguriphase}. At QFT level, OPE provides an useful tool to study their operatorial content \cite{twotwo,Arutyunov:2001hs} while BPS correlators have been helpful in checking the correspondence among DGRT loops on $S^2$ and two-dimensional Yang-Mills \cite{Bassetto:2009rt,Bassetto:2009ms,Giombi:2009ds}. Due to their more complicated geometry, Wilson loop correlators are much more difficult to face, in gauge theory as well as in string theory. When they can be handled, they reveal a rich phenomenology: for example, the correlator of two parallel and concentric circles exhibits at strong-coupling a sort of phase transition, the so-called Gross-Ooguri phase transition \cite{grossooguriphase,gophase1,gophase2}.

Due to the simple geometry, the most studied correlators involve (anti-)parallel lines and circles inside $S^2$. An interesting topologically non-trivial configuration involving great circles is provided by the Hopf fibration of $S^3$: the non-intersecting $S^1$ fibers are in fact ``pierced" together in such a way they cannot be separated. The Hopf fibration has been introduced in the context of Wilson loops in \cite{dgrta}, and it can be naturally considered within the DGRT construction: supersymmetry is also preserved when all the fibers in the same fibration couple the same scalar field with constant coupling. Quite surprisingly the relevant gauge theory propagator between any two points on any two fibers is constant, in Feynman gauge: this suggests that the correlator of any number of Hopf fibers is captured by the same matrix model of the 1/2 BPS circular loop, with different insertion, if interacting diagrams do not contribute. Conversely we would expect that the large $N$, strong-coupling result should be described, at string level, by disconnected world-sheet configurations. While in general there is no reason to believe that interacting diagrams do not contribute, this fact holds true for a single great circle due to its BPS nature: it is natural wondering whether the same simplification occurs also for the BPS configuration involving two Hopf fibers.

In this paper we have tried to give an answer to this question from the string theory point of view, namely we have studied whether a connected minimal area surface connecting two Hopf fibers is subleading with respect to the disconnected solution, or if it does not exist at all. While we have been able to give only a partial answer to this problem, we have found an interesting connected non-BPS string configuration solving the equations of motion with the appropriate boundary conditions. Geometrically, the found solution describes a connected surface linking the Hopf fibers, and it turns out to have smaller area than the disconnected configuration. Moreover, an interesting dynamics resembling the Gross-Ooguri phase transition appears: the connected solution depends on the fiber angular separation in the Hopf base $S^2$ and when the fibers have maximal angular separation, the connected solution degenerates into the disconnected one. As soon as the separation is smaller, the connected solution is distinct from the disconnected one and it has smaller area than the latter. As we will explain in details, the surface describes antiparallel Hopf fibers with the same scalar coupling and hence gives rise to a non-BPS configuration: we will argue that the BPS configuration involving two antiparallel Hopf fibers (with the correct scalar coupling) does not admit any connected string solution, and presumably the parallel fibers neither.

Interestingly enough, the surface associated to the maximal separation is indeed BPS, hence, by moving the fiber angular separation, our solution is able to continuously interpolate between non-BPS and BPS configurations. Moreover, as we will see, the connected world-sheet provides a string description of a system which can be thought of as a deformation of the ordinary antiparallel lines (in the sense explained in the text), representing a sort of deformation of the usual static quark-antiquark potential. We remark that these same topics have been recently discussed in \cite{Drukker:2011za} for the case of the circular loop and antiparallel lines.

The paper is organized as follows. In order to fix some notation, we are going to continue this introduction by briefly reviewing how DGRT Wilson loops are defined, the geometry of the Hopf fibration and the AdS/CFT prescription to compute Wilson loop expectation values, tools we will need for our computations. In sec. \ref{secstringhopf} we will present the derivation of the minimal area surface in $AdS_5\times S^5$ satisfying the desired boundary conditions.
In sec. \ref{secstringsolutions} we will study the properties of the connected solution in the case of constant (no motion in $S^5$) and non-constant (motion in $S^5$) scalar couplings.
In sec. \ref{secSUSY} we will discuss the supersymmetry of the solutions we found. In sec. \ref{secpot} we will explain how the found solution should be interpreted from the gauge theory viewpoint, explaining its relation with the quark-antiquark potential. We will check at perturbative level the consistency of our interpretation. In sec. \ref{secconcl} we will summarize the results, making some comments on the open questions and the directions worth to be studied in the future.

\subsection{The DGRT Wilson loops}\label{secdgtrloops}
The relevant Wilson loop operator in SYM4 is given by \cite{mwloop,pot1,dgo} (in Euclidean signature)
\be 
\begin{split}
&W(\gamma,\theta)\equiv\tr \pexp\left( i\oint_\gamma \de t\;\Big( \dot x^\mu(t)A_\mu(x(t))-i\dot y^I(t)\phi_I(x(t))\Big)\right),\\
&\de y^I=|\dot x(t)|\theta(t)^I\de t\qq \theta^2(t)=1,
\end{split}
\ee
where $A_\mu$ is the gauge connection, $\phi_I$ are the six scalars of the $\N=4$ vector multiplet, $x^\mu$ are $\R^4$ coordinates and $\theta^I(t)$ are $\R^6$ coordinates parameterizing a path in $S^5$ along the loop $\gamma\subset\R^4$ with parameter $t$.

We would like to consider space-time paths lying on some $S^3$ embedded\footnote{$S^3$ can also be seen as a fixed time slice of the compactification $\R\times \R^3\rightarrow \R\times S^3$.}  into $\R^4$. One can construct a generalized gauge connection involving the scalar fields using $SU(2)$ right-invariant 1-forms: in Euclidean coordinates $x^\mu$, constrained by $x^2=1$, they are given by
\footnote{We follow the conventions of \cite{dgrta}.}
\be
\begin{split}
& \sigma^i=2\sigma^i_{\mu\nu}x^\mu\de x^\nu\qq i=1,2,3\qq \mu,\nu=1,2,3,4\\
&\sigma^{k}_{ij}=\ep^{ijk}\qq \sigma^{k}_{i,4}=-\delta^k_i\qq\ep^{123}=1.
\end{split}
\ee
A DGRT Wilson loop is defined by assigning the scalar couplings
\be\label{dgrtscalar} \dot y^I=|\dot x|\theta^I=-\dot x^\mu \sigma^i_{\mu\nu}x^\nu M_i^{\phantom{i}I}\Leftrightarrow \de y^I=\frac{1}{2}\sigma^i M_i^{\phantom{i}I},\ee
where the constant $3\times 6$ matrix $M_{iI}$ is norm preserving\footnote{Because of the $SO(6)$ symmetry, one can consider $M_i^{\ I}=\delta^I_i$.}, i.e. it satisfies
\be M_i^{\phantom{i}I}M_{jI}=\delta_{ij}.\ee
We immediately see that\footnote{For a discussion about the origin of this constraint see for example \cite{dgo}.} $|\dot x|^2=|\dot y^2|$.
As shown in \cite{dgrta}, a DGRT loop is generically 1/16 BPS and supersymmetry is enhanced for particular contours. For example, the familiar 1/2 BPS circular loop, defined by
\be x^\mu=(\cos t,\sin t,0,0)\qq t\in[0,2\pi]\qq \theta^I(t)=\delta^I_3, \ee
belongs to this class, the pullback on the loop of the right-invariant 1-forms being
\be\label{hopfcouplings}\sigma_1=\sigma_2=0\qq \sigma_3=2\de t,\ee
yielding a constant coupling to $\phi_3$ only. Its exact quantum expectation value is known: in Feynman gauge interacting diagrams do not contribute, while the constancy of the combined gauge-scalar propagator on the loop
\be\label{circprop} G^{ab}(x_1,x_2)=\frac{g^2\delta^{ab}}{4\pi^2}\frac{|\dot x_1||\dot x_2|\theta_1^I\theta_{2\,I}-\dot x_1\cdot\dot x_2}{(x_1-x_2)^2}=\frac{g^2\delta^{ab}}{8\pi^2}\ee
allows to sum up all the ladder diagrams, giving the following matrix model representation \cite{eszcircularloop,drukkergrossexact}
\be\label{matrixmodel}
\langle W(S^1)\rangle=\int\frac{\De M}{Z}\;\tr \frac{e^M}{N}\exp\left(-\frac{2N}{\lambda}\tr M^2\right)\qq Z=\int\De M\;\exp\left(-\frac{2N}{\lambda}\tr M^2\right),
\ee
where $M$ is a constant $N\times N$ Hermitian matrix. In particular, in the large $N$, $\lambda$ limit one has
\be \langle W(S^1)\rangle \simeq\sqrt{\frac{2}{\pi}} \lambda^{-3/4} e^{\sqrt{\lambda}} .\ee
The same matrix model has been obtained in \cite{matrixmodelloc} by formulating the theory on $S^4$ and applying a localization procedure to compute exactly the path-integral: actually, in the same paper, a much more general matrix integral has been derived for ${\cal N}=2$ superconformal theories, including non-trivial instanton corrections. 
\subsection{The Hopf fibration}\label{sechopf}
From the geometrical point of view $S^3$ can be described as a non-trivial $S^1$ bundle over $S^2$: the construction is known as the Hopf fibration. Its structure becomes transparent by parameterizing $S^3\subset\R^4$ by\footnote{We will refer to these angles as Hopf coordinates.} $(\theta,\phi,\psi)\in[0,\pi]\times[0,2\pi]\times[0,4\pi]$
\be
\label{hopfcoord}
\begin{split}
&x^1=-\sin\frac{\theta}{2}\sin\frac{\psi-\phi}{2}\qq x^2=\sin\frac{\theta}{2}\cos\frac{\psi-\phi}{2},\\
&x^3=\cos\frac{\theta}{2}\sin\frac{\psi+\phi}{2}\qq x^4=\cos\frac{\theta}{2}\cos\frac{\psi+\phi}{2}
\end{split}
\ee
and projecting with the Hopf map $\pi$
\be\pi:\left(\begin{array}{c}x^1\\x^2\\x^3\\x^4\end{array}\right)\in
S^3\mapsto\left(\begin{array}{l}2(x^2x^4-x^1x^3)\\
2(x^2x^3+x^1x^4)\\
x_3^2+x_4^2-x_1^2-x_2^2\end{array}\right)=
\left(\begin{array}{l}\cos\phi\sin\theta\\
\sin\phi\sin\theta\\
\cos\theta\end{array}\right)\in S^2.\ee
The projection reveals that $(\theta,\phi)$ parameterize the base $S^2$ while $\psi$ the fibers $S^1$, which are non-intersecting great circles inside $S^3$.
\begin{figure}[!h]
\centering\includegraphics{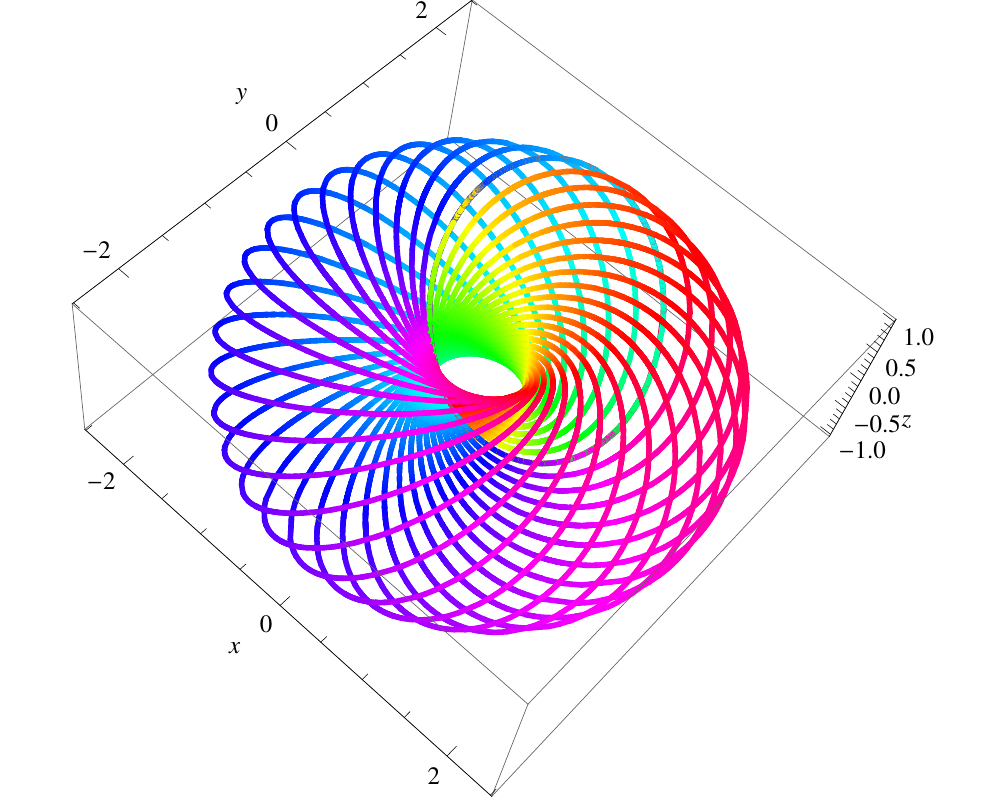}
\caption{\footnotesize The stereographic projection of $S^3$ onto $\R^3$ representing the Hopf fibration. For simplicity, in the picture are shown the fibers at $\phi=\rm const.$, lying on a torus. Points of the same color correspond to the same value of $\psi$.}
\end{figure}

The Hopf fibration has been introduced in the context of supersymmetric Wilson loops in \cite{dgrta}. By expressing the right invariant 1-forms in Hopf coordinates
\be
\begin{split}
\sigma_1=&-\sin\psi\de\theta+\cos\psi\sin\theta\de\phi,\\
\sigma_2=&\cos\psi\de\theta+\sin\psi\sin\theta\de\phi,\\
\sigma_3=&\de\psi+\cos\theta\de\phi,
\end{split}
\ee
we can see that all the fibers in the same fibration couple to the same scalar. In fact, at constant $(\theta,\phi)$ we have
\be\sigma_1=\sigma_2=0\qq \sigma_3=\de\psi\Rightarrow \sigma_3=2\de t\qq \psi(t)=2t,\;t\in[0,2\pi]\ee as in \reff{hopfcouplings}. Remarkably 
the loops share half the supersymmetries of a single circle, the system being effectively 1/4 BPS. The corresponding preserved supercharges will be essentially the same as the ones associated to the 1/2 BPS maximal circle, except that only one chirality is selected \cite{dgrta}. It is therefore natural to expect that correlators of Hopf fibers should be exactly computable, due to their BPS character. At perturbative level we observe another peculiar feature pointing in this direction, namely that in a configuration made of Hopf fibers, the gauge-scalar propagator (in Feynman gauge) between any pair of points $x_1$, $x_2$ on any fiber is the same constant as in \reff{circprop}. As pointed out in \cite{dgrta}, this implies that all the ladder diagrams contributing to the correlator of $n$ fibers are summed up by the same matrix model of the 1/2 BPS circle \reff{matrixmodel}, but with a different insertion
\be\langle W(n)\rangle_\text{ladder}=\int\frac{\De M}{Z}\;\left(\frac{\tr e^M}{N}\right)^n\exp\left(-\frac{2N}{\lambda}\tr M^2\right).\ee
As mentioned in the beginning, in the large $N$ limit the ladder correlator will be the same as $n$ non-interacting circles, and it will be reproduced at strong coupling by $n$ disconnected string surfaces. The absence of dominant connected configurations would be a stringy support to the conjecture about the exactness of the matrix model\footnote{This conjecture is also supported by the cancellation of the lowest order diagrams containing interactions  \cite{PCom}.} for the Hopf fibration \cite{dgrta}.

\subsection{The string description of Wilson loops in ${\cal N}=4$ SYM}\label{secstring}
The AdS/CFT prescription to compute the expectation value of Wilson loops, originally proposed in \cite{mwloop,pot1}, can be schematically described as
\be\label{stringwm} \langle W(\mathcal{C})\rangle =
Z_\textrm{string}[\mathcal{C}]\equiv\int_{\pa X=\mathcal{C}}\De X
e^{-\sqrt{\lambda}S[X]}\qq \mathcal{C}\equiv(\gamma;\theta),\ee where
$S$ is the string action in the $AdS_5\times S^5$ background and $X$ represents all the string coordinates.
The sum over the string world-sheets is taken over those ending on the loop $\gamma\subset\pa AdS_5$ and lying along
$\theta\in S^5$ (at the AdS boundary). These boundary conditions arise in a natural way by considering the 10-dimensional nature of SYM$_{4}$ (see \cite{dgo} for a detailed derivation).

The string action is given by a non-linear sigma model\footnote{The original derivation of the full action can be found in \cite{adsstring}. We are interested just in the leading semiclassical approximation of the path-integral, where the fermions can be taken vanishing. We display therefore only the bosonic sector of the full action.}, namely
\be S[X]=\frac{1}{4\pi}\int\de^2\sigma\sqrt{\gamma}\gamma^{ab} G_{MN}(X)\pa_a X^M\pa_b X^N\qq X^M\in AdS_5\times S^5,\ee
where $\sigma^a$ are world-sheet coordinates, $\gamma_{ab}$ is the world-sheet metric and $G_{MN}(X)$ is the $AdS_5\times S^5$ metric: the action is classically equivalent to the area functional. Using the $\gamma_{ab}$ equations, that is the vanishing of the world-sheet energy-momentum tensor
\be  0=-4\pi\frac{1}{\sqrt{\gamma}}\frac{\delta S}{\delta
\gamma^{ab}}\equiv T_{ab}=h_{ab}
-\frac{1}{2}\gamma_{ab}\gamma^{cd}h_{cd}\qq h_{ab}\equiv G_{MN}(X)\pa_a X^M\pa_b X^N, \ee
the action becomes
\be S[X]=\frac{1}{2\pi}A[X]=\frac{1}{2\pi}\int\de^2\sigma\sqrt{h},\ee
where $h_{ab}$ denotes the induced metric on the world-sheet.
In the large $N$ limit, the expectation value of a Wilson loop at strong coupling is computed as a classical solution of the string equations (i.e. the minimal area surface) in $AdS_5\times S^5$ with the boundary conditions  imposed by the loop and the scalar couplings
\be \langle W(\gamma,\theta)\rangle\simeq e^{-\sqrt{\lambda}S_\text{min}[\gamma,\theta]}.\ee

In the next section we are finally going to face the problem of finding a minimal string surface ending on a pair of Hopf fibers. It is worth to outline the strategy:
\begin{itemize}
\item[-] We will begin by writing the $AdS_5\times S^5$ sigma model employing a coordinate system adapted to the Hopf fibration \reff{hopfcoord}.
\item[-] We will give a natural ansatz based on the Hopf fibration and $AdS_5\times S^5$ geometries. This allows us to simplify the problem.
\item[-] We will verify that the ansatz gives a consistent solution, and we will discuss its properties.
\end{itemize} 

 \section{The string description of Hopf fiber correlators}
 \label{secstringhopf}
We will work in the conformal gauge, namely we will consider world-sheet coordinates such that $\sqrt{\gamma}\gamma^{ab}=\delta^{ab}$ (Euclidean signature). Then the equations $T_{ab}=0$ must be imposed as constraints (Virasoro constraints)
\be
\begin{split}
\text{I.}\ \ \ & T_{11}=\frac{1}{2}\left(h_{11}-h_{22}\right)=-T_{22}=0,\\
\text{II.} \ \ \ & T_{12}=h_{12}=0,
\end{split}
\ee
while the $X^M$ equations describe a geodesic-like motion\footnote{We mean that the lines $\sigma^1=const.$ or $\sigma^2=const.$ are geodesics of the target space.} in $AdS_5\times S^5$. Thanks to the product geometry of the target space, coordinates where $G_{MN}$ is diagonal with respect to the factor spaces can be chosen, namely
\be G_{MN}(X)=\left(\begin{array}{cc}G^{AdS}_{mn}(\rho)&0\\0&G^{S}_{IJ}(\Theta)\end{array}\right),\ee
where $\rho^m$, $G^{AdS}_{mn}$ are coordinates and metric on $AdS_5$, and $\Theta^I$, $G^{S}_{IJ}$ are coordinates and metric on $S^5$. The world-sheet energy-momentum tensor and the induced metric receive separate contributions from the two factors, ending up with two distinct sigma models
\be
\begin{split}
& S=S_{AdS}+S_{S}\qq S_{AdS}=\frac{1}{4\pi}\int\de^2\sigma\,\mathcal{L}^{AdS}\qq S_{S}=\frac{1}{4\pi}\int\de^2\sigma\,\mathcal{L}^{S}\\
&
\mathcal{L}^{AdS}=h_{11}^{AdS}+h_{22}^{AdS}\qq\mathcal{L}^{S}=h_{11}^{S}+h_{22}^{S},
\end{split}
\ee which must be solved with the following boundary conditions (schematically)
\be \rho|_{\pa AdS}=\gamma\qq \Theta|_{\pa AdS}=\theta.\ee
However, the systems are still interacting through the Virasoro constraints
\be 0=T_{ab}=T^{AdS}_{ab}+T^{S}_{ab}\Rightarrow T^{AdS}_{ab}=-T^{S}_{ab}\ee
and through the world-sheet coordinates, which must be the same on both the factor spaces.

\subsection{$AdS_5$ motion}

We will work with $AdS_5$ in global coordinates
\be
\begin{split}
&\tau\qq\rho\qq\Omega^\mu\qq\mu=1,2,3,4\\
&\tau\in(-\infty,+\infty)\qq \rho\in[0,+\infty)\qq \Omega^2=1,
\end{split}
\ee
where the metric takes the form
\be\de s^2_{AdS}=\cosh^2\rho\de\tau^2+\de\rho^2+\sinh^2\rho\de\Omega_{3}^2.\ee
In this coordinate system the AdS boundary is mapped to $\rho\to\infty$. Moreover, since the 3-sphere at infinity parametrized by $\Omega^\mu$ is naturally
identified with a fixed time slice of the global AdS boundary $\R\times S^3$, we will employ a coordinate system adapted to the Hopf fibration \reff{hopfcoord}
\be
\begin{split}
&\Omega^1=-\sin\frac{\theta}{2}\sin\frac{\psi-\phi}{2}\qq\Omega^2=\sin\frac{\theta}{2}\cos\frac{\psi-\phi}{2}\\
&\Omega^3=\cos\frac{\theta}{2}\sin\frac{\psi+\phi}{2}\qq\Omega^4=\cos\frac{\theta}{2}\cos\frac{\psi+\phi}{2}.
\end{split}
\ee The $S^3$ metric $\de \Omega_3^2$ reads as
\be \de\Omega^2_3=\frac{1}{4}\left(\de\theta^2+\de\phi^2+\de\psi^2+2\cos\theta\de\phi\de\psi\right),\ee
so that the $AdS_5$ metric with respect to $(\rho^m;m=1,\ldots5)\equiv(\tau,\rho,\theta,\phi,\psi)$ is given by
\be
G^{AdS}=\left(\begin{array}{cc|ccc}
\cosh^2\rho& & & &\\
 & 1& & &\\ \hline
 &  &\frac{\sinh^2\rho}{4}\left(\begin{array}{ccc}1&&\\&1&\cos\theta\\&\cos\theta&1\end{array}\right)\end{array}\right).\ee
We will take the world-sheet coordinates to be $\sigma^a\equiv(t,s)$; differentiation with respect to $t$ and $s$ will be denoted by a dot and a prime respectively. The form of the $AdS_5$ metric suggests the following notation
\be
 |v|^2_{S^3}\equiv g_{mn}v^m v^n\qq g_{mn}\equiv \frac{G^{AdS}_{mn}}{\sinh^2\rho}\qq m,n=3,4,5
\ee
 $g_{mn}$ representing the metric tensor
 \footnote{$g_{mn}$ transforms as a tensor with respect to coordinate changes of the form  
 \[
 (\tau,\rho,\Omega)\mapsto(\tau'=\tau'(\tau,\rho),\rho'=\rho,\Omega'=\Omega'(\Omega))
 \]
  In particular, it is a tensor with respect to reparameterizations of $S^3$.}
  of $S^3$. By setting $v\equiv(\theta,\phi,\psi)$, the induced metric $h^{AdS}$ is explicitly given by
\be
\begin{split}
&h_{11}^{AdS}=\dot\tau^2\cosh^2\rho+\dot\rho^2+|\dot v|^2_{S^3}\sinh^2\rho,\\
&h_{12}^{AdS}=\dot\tau\tau'\cosh^2\rho+\dot\rho\rho'+\frac{\sinh^2\rho}{4}\left(\dot\theta\theta'+\phi'(\dot\phi+\dot\psi\cos\theta)+\psi'(\dot\psi+\dot\phi\cos\theta)\right),\\
&h_{22}^{AdS}=\tau'^2\cosh^2\rho+\rho'^2+|v'|^2_{S^3}\sinh^2\rho.
\end{split}
\ee

\subsection{$S^5$ motion} We begin by considering the 5-sphere parametrized by $\Theta\in\R^6$, $\Theta^2=1$. As far as DGRT loops
are concerned, we can limit  to consider a non-trivial motion in a
subspace $S^2\subset S^5$ \cite{dgrta}. By setting
$\Theta^4=\Theta^5=\Theta^6=0$ and parameterizing the relevant sphere $S^2\subset S^5$ as \be
\Theta_1=\cos\beta,\qq \Theta_2=\cos\alpha\sin\beta,\qq \Theta_3=\sin\alpha\sin\beta,\ee
the $S^2\subset S^5$ metric is given by
\be G^{S}=\left(\begin{array}{cc}1&\\&\sin^2\beta\end{array}\right)\ee
and the induced metric $h^S$ reads as 
\be
\begin{split}
& h_{11}^S=\dot\beta^2+\sin^2\beta\dot\alpha^2=|\dot\xi|^2_{S^2},\\
& h_{12}^S=\dot\beta\beta'+\sin^2\beta\dot\alpha\alpha',\\
&h_{22}^S=\beta'^2+\sin\beta\alpha'^2=|\xi'|^2_{S^2},
\end{split}
\ee 
where we have set $(\xi^i;i=1,2)\equiv(\beta,\alpha)$.

\subsection{$AdS_5$ ansatz and boundary conditions}
In order to make manageable the sigma model equations, we tailor an ansatz inspired by the Hopf fibration and $AdS_5\times S^5$ geometries. In this way we will be able to find a classical solution to the equations of motion and verify that the boundary conditions are satisfied. We will consider separate ansatz for the $AdS_5$ and $S^5$ motions, examining any compatibility conditions later.

In the given coordinate system, the variable $\psi$ describes a Hopf fiber, and we can consider $\psi$ as the loop parameter on the AdS boundary. Since each point of the base $S^2$ determines a unique
fiber, we look for a solution having the asymptotic form
\be
\begin{split}
&\rho\to\infty\qq\tau=\mathrm{const.},\\
&\theta=\mathrm{const.}\qq\phi=\mathrm{const.}\qq\psi(t)=2kt\qq t\in[0,2\pi],
\end{split}
\ee
 where $k\in \mathds{Z}$ is a constant denoting the winding
number of the loop around the fiber\footnote{For us $k=\pm1$. However, we are not going to specify it.}. We recognize a situation similar to the 1/2 BPS circular loop or to the correlator of parallel circles, where one of the world-sheet parameters ($t$) is taken to be the angular coordinate describing the circle(s). In those cases, the surfaces are simply obtained by taking the radial coordinate $\rho$ as a function of the other parameter ($s$) and allowing an $s$-dependence only for the global time $\tau$. However, in the present case this certainly cannot work since we would like to find a connected solution linking two fibers in the same fibration. Since two fibers are surely at different points of the base $S^2$, we let the base angle $\theta$ and $\phi$ depend on the surface parameters. For this reason, our $AdS_5$ ansatz is
\be
\label{adsansatz}
\begin{split}
& \tau=\tau(s)\qq \rho=\rho(s)\qq
\theta=\theta(s)\qq\phi=\phi(s)\\
&\psi=\psi(t,s)=2kt+\eta(s)\qq t\in[0,2\pi]\qq s\in[s_1,s_2]
\end{split}
\ee
 where we have allowed $\psi$ to have
a simple $s$-dependence to treat the $S^3$ angles on equal footing. The ansatz above must be considered together with the boundary conditions \be
\lim_{s\to s_{1,2}}\rho(s)=+\infty\qq
\theta(s_{1,2})=\theta_{1,2}\qq\phi(s_{1,2})=\phi_{1,2}\ee
where $(\theta_1,\phi_1)$ and $(\theta_2,\phi_2)$ represent the two fibers over the respective points of the base $S^2$.

As far as Wilson loop correlators are concerned, a typical feature is
the existence of a turning point for the extension of the
surface into the AdS interior, namely a minimal value $\rho_0>0$
for the radial coordinate. In our case, the radial coordinate is a function of one parameter only ($s$), so that the turning points are determined by the minima of $\rho(s)$
\be \rho'(s_0)=0\qq \rho_0\equiv\rho(s_0)\ee We expect that the connected surface will start
on one fiber in the AdS boundary at $s=s_1$, reaching a turning point at
$s=s_0$, and then will come back towards the boundary on the other fiber at
$s=s_2$.

\subsection{$S^5$ ansatz and scalar couplings}

As previously stated, for DGRT loops we can  limit  to consider a motion in $S^2\subset S^5$ (in fact, at most three scalars are turned on, parameterizing a 2-sphere). In the following, in order to avoid possible confusion between the Hopf base $S^2$ and $S^2\subset S^5$, we will often refer to the latter 2-sphere simply as $S^5$, having in mind that the dynamics is actually developing in the $S^2\subset S^5$ subspace. Since the scalar couplings are given by \be
\sigma_1=\sigma_2=0\qq \sigma_3=\de\psi\Rightarrow \sigma_3=2k\de t,\ee the $S^5$ parameters $\theta^I$ in the gauge theory are the same constant for parallel fibers or differ by the sign for antiparallel ones
\be\label{scalarcharge}\frac{\dot y^I}{|\dot y|}=\theta^I(t)=\textrm{sign}(k)\delta^I_3.\ee
For parallel fibers the string surface can sit entirely at one point of $S^5$. In this case the $S^5$ motion is trivially solved by constant values of $\alpha$, $\beta$. Instead, antiparallel fibers sit at antipodal points of $S^5$ at the AdS boundary, and a motion in $S^5$ joining them is needed. We will consider the simplest situation, namely a geodesics motion in the parameter $s$. Without loss of generality we can set\footnote{This solution is analogous to the $S^1$ ansatz considered in \cite{drukkerfiol}.}
\be\label{s5motion} \alpha(s)=\alpha_0\qq\beta(s)=\beta_0+c(s-s_0)\qq c=const.\in\R,\ee
where $\alpha_0$, $\beta_0$, $s_0$ are constants determining the boundary values of the $S^5$
angles, and $|\xi'|^2_{S^5}=c^2$. We can see that the $T^S$ components are constant and given by
\be T^S_{11}=-\frac{c^2}{2}\qq T^S_{12}=0,\ee
whereas $c^2=0$ when there is no $S^5$ motion.

\subsection{General analysis of the motion}
By plugging the $AdS_5\times S^5$ ansatz into the induced metric $h=h^{AdS}+h^{S}$, we get
\be
\label{inducedmetric} 
\begin{split}
&h_{11}=k^2\sinh^2\rho,\\
& h_{12}=\frac{k}{2}\sinh^2\rho\left(\eta'+\phi'\cos\theta\right),\\
&h_{22}=\rho'^2+|v'|^2_{S^3}\sinh^2\rho+\tau'^2\cosh^2\rho+c^2,
\end{split}
\ee
so that the sigma model is now reduced to a 1-dimensional system. Since $\pa_\tau\mathcal{L}^{AdS}=\pa_\phi\mathcal{L}^{AdS}=\pa_\eta\mathcal{L}^{AdS}=0$,
we can easily obtain three constants of motion and a first integration of the $\tau,\phi,\eta$ equations\footnote{We will denote by $c_{\ldots}$ a generic integration constant.}
\be
\begin{split}
\label{hopfgravity07}\tau&:\cosh^2\rho\tau'=c_\tau\Rightarrow\tau'=\frac{c_\tau}{\cosh^2\rho},\\
\phi&:\sinh^2\rho(\phi'+\cos\theta\eta')=c_\phi,\\
\eta&:\sinh^2\rho(\eta'+\cos\theta\phi')=c_\eta.
\end{split}
\ee
Moreover, as it can be inferred from the form of $\mathcal{L}^{AdS}$
\be \mathcal{L}^{AdS}=\rho'^2+\sinh^2\rho(k^2+|v'|^2_{S^3})+\tau'^2\cosh^2\rho,\ee
the variables $(\theta,\phi,\eta)$ describe a geodesic motion in $S^3$ in the parameterization
\be\label{veq} |v'|_{S^3}\propto \frac{1}{\sinh^2\rho}\ee
namely\footnote{$
\Gamma^\phi_{\theta\phi}(S^3)=\Gamma^\psi_{\theta\psi}(S^3)=\frac{1}{2}\frac{\cos\theta}{\sin\theta}\qq
\Gamma^\phi_{\theta\psi}(S^3)=\Gamma^\psi_{\theta\phi}(S^3)=-\frac{1}{2\sin\theta}\qq
\Gamma^\theta_{\phi\psi}(S^3)=\frac{1}{2}\sin\theta.$}
\be
v''^m+\Gamma^m_{nk}(S^3)v'^n v'^k= v'^m\pa_s\ln\frac{1}{\sinh^2\rho}\ee and then we can set
\be |v'|^2_{S^3}=\frac{c^2_v}{\sinh^4\rho}.\ee
Therefore it is useful to work out explicitly only the $\rho$ equation
\be\rho:\rho''=\sinh\rho\cosh\rho\left(k^2+|v'|^2_{S^3}+\tau'^2\right).\ee

Now we have to verify the compatibility of the ansatz with the
Virasoro constraints: the first one is $T^{AdS}_{11}=c^2/2$,
which reads  \be
 2T^{AdS}_{11}=\sinh^2\rho\left(k^2-|v'|^2_{S^3}\right)-\rho'^2-\cosh^2\rho\tau'^2=c^2.\ee
It encodes almost all the dynamical information\footnote{In fact, it is the vanishing of the ``energy'': $E=\rho'^2-\sinh^2\rho\left(k^2-|v'|^2_{S^3}\right)+\tau'^2\cosh^2\rho+c^2$.}. By differentiating $T^{AdS}_{11}=c^2/2$ with respect to $s$ we get 
\be
\begin{split}
0&=\rho'\left(\sinh\rho\cosh\rho\left(k^2+|v'|_{S^3}^2+\tau'^2\right)-\rho''\right)+\\
&-\left(\frac{1}{2}\sinh^2\rho\pa_s|v'|_{S^3}^2+|v'|_{S^3}^2\pa_s\sinh^2\rho\right),
\end{split}
\ee
where we have used the $\tau$ equation. The $\rho$
equation requires the vanishing of the second term, namely
\be \pa_s(\sinh^2\rho|v'|_{S^3})=0\Leftrightarrow|v'|_{S^3}=\frac{c_v}{\sinh^2\rho}\qq c_v>0\ee
which is compatible with the geodesic motion in $S^3$ previously described. At this point we can see how an arbitrary $S^5$ motion could spoil our $AdS_5$ ansatz: $\pa_s T^S_{11}\neq 0$ would be
incompatible with $|v'|_{S^3}\propto 1/\sinh^2\rho$. The second constraint $T^{AdS}_{12}=0$ reads instead
\be
T^{AdS}_{12}=\frac{k}{2}\sinh^2\rho\left(\eta'+\cos\theta\phi'\right)=0,\ee
which simply states that the $\eta$ equation \reff{hopfgravity07} is satisfied with $c_\eta=0$. In order to have $\rho\not\equiv0$ we must impose
\be\label{secondvir} \eta'+\cos\theta\phi'=0\ee
and hence the dynamics can be entirely expressed in terms of the base angles. Actually, only one angle is dynamically independent since \reff{hopfgravity07}, \reff{veq}, \reff{secondvir} yield
\be\label{phietaeq} \phi'=\frac{c_\phi}{\sin^2\theta\sinh^2\rho}\qq |v'|^2_{S^3}=\frac{1}{4}\left(\theta'^2+\phi'^2\sin^2\theta\right)=\frac{c_v^2}{\sinh^4\rho},\ee \be\label{thetaeqbis}\theta'^2=\frac{4c_v^2}{\sin^2\theta\sinh^4\rho}\left(\sin^2\theta-\frac{c_\phi^2}{4c_v^2}\right).\ee
We observe that the second Virasoro constraint acts as a projection on the base, and then $(\theta,\phi)$ describe geodesics on a 2-sphere.

\section{String solutions}
\label{secstringsolutions}
\subsection{Single fiber: 1/2 BPS circular loop} As a byproduct of the previous analysis, we can derive the well known result \cite{global12bps} of the 1/2 BPS circular loop, which is simply the case of a single Hopf fiber. This result will be useful in the following.

As remarked in the previous section, constant scalar couplings allows to consider
a surface sitting entirely at one point of $S^5$ (at least in the
case of a single loop), and in our case this translates into
$c^2=0$. Let us consider $\tau'=0$, that is $c_\tau=0$ (i.e. the
surface will lay at a constant global time slice). Since we do not
need a surface connecting different fibers, we set also $c_v=0$,
i.e. $\theta'=\phi'=\eta'=0$. The $\rho$ equation reads therefore \be
\rho'^2=k^2\sinh^2\rho\Rightarrow\rho'=\pm |k| \sinh\rho,\ee and the
induced metric is given by \be
h_{ab}=\left(\begin{array}{cc}k^2\sinh^2\rho&\\&\rho'^2\end{array}\right).\ee
Since $\rho'(s)$ has a definite sign in each (equivalent) branch, we
can consider $\rho$ as a new parameter and the well known
$AdS_2$ geometry is manifestly recovered ($k^2=1$)\be\label{rho12}
\de s^2=\sinh^2\rho\de t^2+\de\rho^2.\ee The function $\rho(s)$ is computed 
easily \be \rho(s)=2\tanh^{-1}\left(e^{\pm
(s-s_\infty)}\right),\ee but we do not need its explicit form to
determine the area, which is given by
\be
\begin{split}
 A_1=&\int\de t\de s\,\sqrt{h}=\int\de t\de s\,\rho'\sinh\rho\\
=&\int_0^{2\pi}\de t\int_0^{1/\ep'}\de\rho\sinh\rho=2\pi \left(\cosh\frac{1}{\ep'}-1\right)\qq \ep'\to 0,
\end{split}
\ee where we have
considered the usual cutoff $\ep'$ to regularize the boundary divergence. The regularized and finite area is\footnote{Instead of the area functional, we should consider a Legendre transform \cite{dgo}. It can verified that the two results agree.}
\be A_1^R=-2\pi. \ee

The single circle solution is extended immediately to a solution describing a set of $n$ independent minimal surfaces by simply letting the fiber constants $(\theta,\phi)$ to
belong to a set of distinct $n$ pairs $\{(\theta_{i},\phi_{i});i=1,\ldots,n\}$: this is the string
description of the disconnected correlator of $n$ Hopf fibers. In particular, the disconnected correlator of two Hopf
fibers is described by two copies of the surface found above, whose area will be $2 A^R_1=-4\pi $.

\subsection{Two fibers: Hopf fibers correlator}
In order to find the connected contribution to the correlator of two
Hopf fibers, we have to allow a smooth motion between two different
fibers. Therefore we will take\footnote{We are rescaling all the $c$
constants by $|k|$.}
\be\label{vpeq}|v'|_{S^3}=\frac{|k|p}{\sinh^2\rho}\qq |k|p\equiv
c_v>0.\ee Keeping\footnote{It can be verified that, given the relation between global and
Poincar\'e coordinates (see appendix \ref{appint}), $\tau=0$
corresponds to the unit 3-sphere in $\R^4$.} $\tau=0$, the $\rho$
equation is 
\be
\begin{split}
\rho'^2&=k^2\sinh^2\rho\left(1-\frac{p^2}{\sinh^4\rho}\right)-k^2c^2=
\frac{k^2}{\sinh^2\rho}\left(\sinh^4\rho-c^2\sinh^2\rho-p^2\right),\\
\rho''&=k^2\sinh\rho\cosh\rho\left(1+\frac{p^2}{\sinh^4\rho}\right).
\end{split}
\ee
We can see that $\rho'$ has two branches of opposite sign separated by a turning point: since $\rho''>0$, the zeros of $\rho'^2$ are minima. Actually there is a unique turning point $\rho_0$ given by
\be\label{turningpoints}\sinh\rho_0=\sqrt{\frac{c^2}{2}+\sqrt{\frac{c^4}{4}+p^2}}.\ee

A solution of this type is candidate to describe the
Hopf fiber correlator: let us consider the two possible situations on $S^5$.

\subsection{No motion in $S^5$}
Setting $c=0$ the surface will lay at a single point on $S^5$. The
$\rho$ equation simplifies to \be\label{rhoequations}
\rho'^2=\frac{k^2}{\sinh^2\rho}\left(\sinh^4\rho-p^2\right)\Rightarrow\rho'=
\pm\frac{|k|}{\sinh\rho}\sqrt{\sinh^4\rho-p^2}\ee and the turning
point separating the two branches is determined by \be
\sinh\rho_0=\sqrt{p}\quad\text{ or }\quad \cosh\rho_0=\sqrt{1+p}.\ee
Since $\rho$ is a monotonic function of $s$ in each branch, we can
express the induced metric in terms of $t$ and $\rho$ as we did for
the single fiber\footnote{The turning point is a coordinate
singularity.} \be \de s^2=k^2\sinh^2\rho\de
t^2+\frac{\sinh^4\rho}{\sinh^4\rho-p^2}\de\rho^2.\ee The area of the
surface is given by \be\label{area2} A_2(p)=2\int_0^{2\pi}\de
t\int_{\rho_0}^\infty\de\rho\sqrt{h}=4\pi
|k|\int_{\rho_0}^\infty\de\rho\,\frac{\sinh^3\rho}{\sqrt{\sinh^4\rho-p^2}}\ee
\begin{figure}
{\includegraphics[width=7cm]{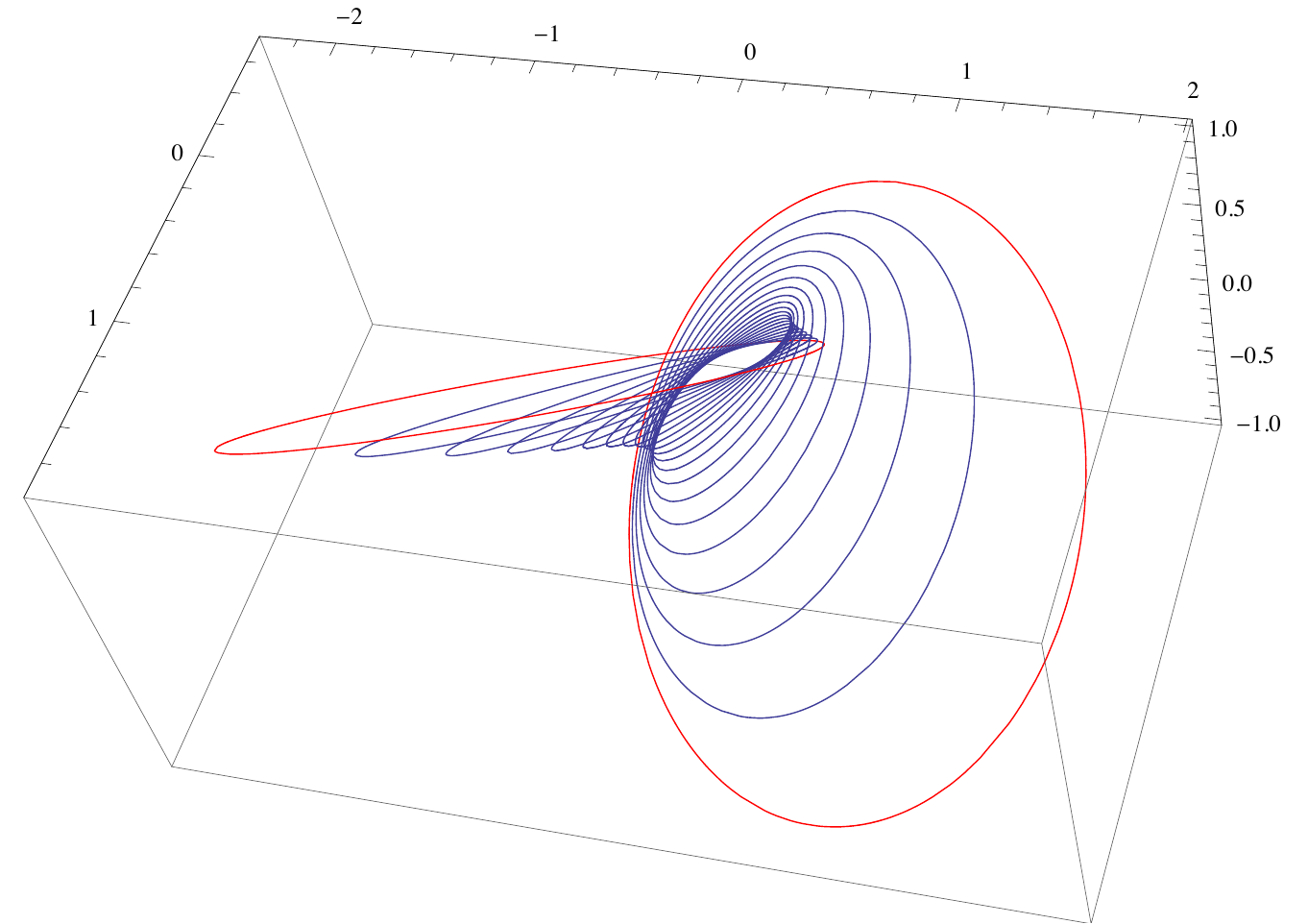}}
{\includegraphics[width=7cm]{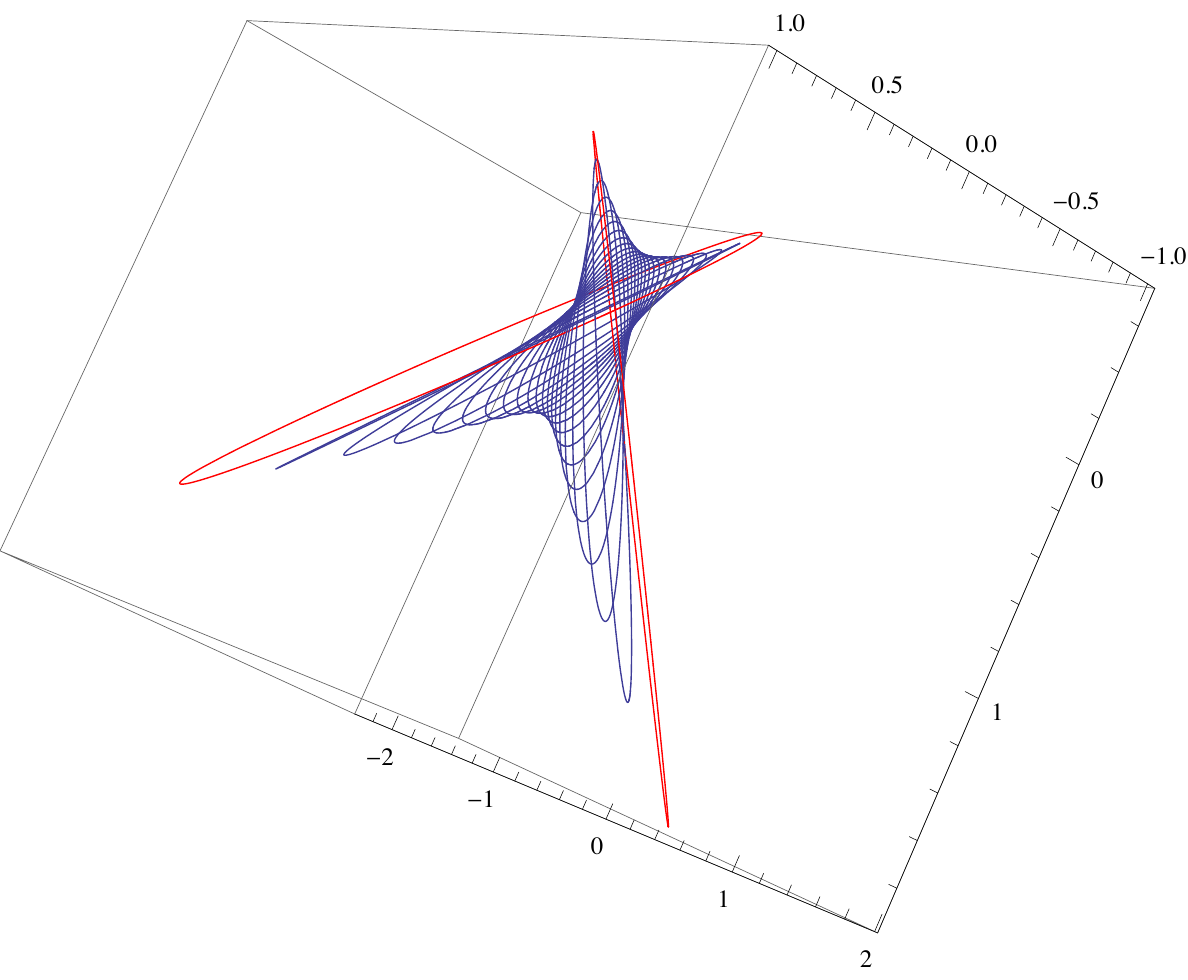}} \caption{\footnotesize A
representation of the connected surface by means of the
stereographic projection of $S^3$. The fixed global time slice of
$AdS_5$ is foliated by a family of 3-spheres of radius $\sinh\rho$.
The red circles represent the two boundary fibers at $\rho=\infty$,
while the blue ones correspond to fibers going into the AdS
interior. The fiber radius decreases by moving in the radial
direction from the boundary up to the turning point, where the
radius of the 3-sphere is $\sinh\rho_0=\sqrt{p}$.}\label{figconnsol}
\end{figure}where a factor of 2 has been considered because of the two branches. By setting
\be x\equiv\cosh\rho\qq b\equiv\sqrt{1+p}\qq
a\equiv\sqrt{\pm(p-1)}\ee the area is given by the integral
\be
\label{area2int}
\begin{split} 
\frac{A_2(p)}{4\pi|k|}=&\lim_{u\to\infty}\int_{\sqrt{1+p}}^{u}\de x\frac{x^2-1}{\sqrt{(x^2-1)^2-p^2}}\\
=&\lim_{u\to\infty}\int_{b}^{u}\de x\frac{x^2-1}{\sqrt{(x^2-b^2)(x^2\pm a^2)}}=I_1(p)+u,
\end{split}
\ee 
where the limit $u=\cosh(1/\ep')\to\infty$ is understood, and we have defined\footnote{The elliptic integrals of the first, second and third kind with argument $\phi$, modulus $m$ and parameter $n$ are denoted by $F(\phi|m)$, $E(\phi|m)$, $\Pi(\phi,n|m)$ respectively. When $\phi=\frac{\pi}{2}$ the argument is omitted and $F(m)$ is denoted by $K(m)$. To agree with the conventions of \cite{int}, we should replace $m\to\sqrt{m}$ and exchange $\phi\leftrightarrow n$.}
\be\label{area2ell} I_1(p)\equiv\left\{\begin{array}{ll}\frac{b^2-1}{b}K(\frac{a^2}{b^2})-b E(\frac{a^2}{b^2})&\textrm{ for } a^2=1-p>0\\
 & \\
\frac{b^2-1}{\sqrt{a^2+b^2}}K(\frac{a^2}{a^2+b^2})-\sqrt{a^2+b^2}E(\frac{a^2}{a^2+b^2})&\textrm{
for } a^2=p-1>0.\end{array}\right.\ee
Since these integrals are finite for every allowed value of $p$, in this form the
divergence coming from $u\to\infty$ is manifest. However, it is
exactly the area divergence of the disconnected solution and then
the difference \be\label{area2reg} \frac{\Delta A(p)}{4\pi |k|} \equiv\frac{A_2(p)-2A_1}{4\pi
|k|}=I_1(p)+1\ee is well
defined\footnote{The same result is obtained by performing the
Legendre transform.} for every fixed $p$.
\begin{figure}[!h]
\centering\includegraphics[width=8cm]{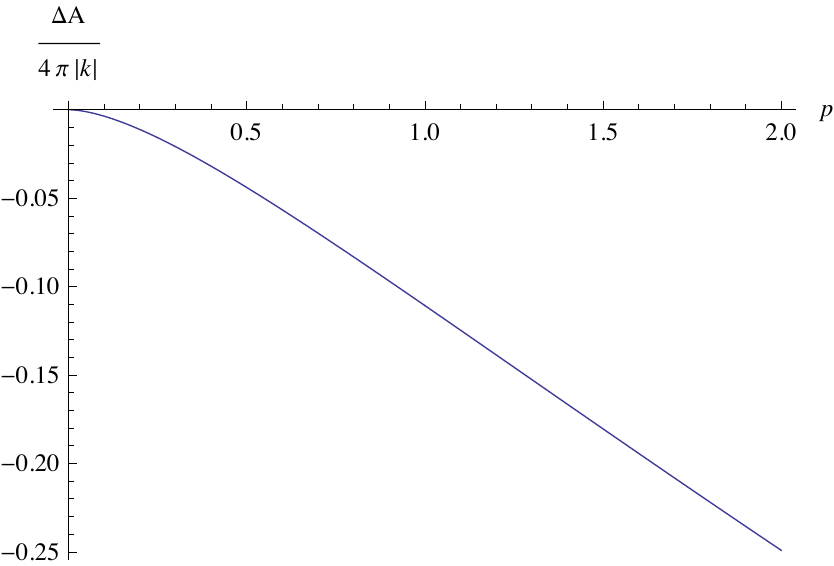} \caption{\footnotesize
Plot of $\frac{\Delta A}{4\pi|k|}$ as a function of
$p$.}\label{figDAcorr}
\end{figure}From the plot shown in Fig.\ref{figDAcorr}, we can see that the
connected surface has smaller area than the disconnected one for any $p$ ($\Delta A(p)\leq 0$): the connected solution always dominates. For $p\to\infty$, $\Delta A(p)$ diverges: as shown below, this corresponds to the limit of coincident fibers.

From \reff{vpeq} we can see that the parameter $p$ is related to the boundary conditions on the angular coordinates of
$S^3$. In order to make everything as simple as possible\footnote{Actually, the general case is straightforward. In fact, upon the use of the second Virasoro constraint, the parameter $p$ can be related to the geodesic (or Euclidean) distance of the fibers on the base, which can be expressed in terms of $(\theta_i,\phi_i)$.}, we can consider the particular solution with $\phi'=0$ (and hence also $\eta'=0$). In this case each fiber is identified uniquely by
$\theta$, which is in turn determined by (assuming $\theta'>0$)
\be |v'|_{S^3}=\frac{|k|p}{\sinh^2\rho}=\frac{\theta'}{2}.\ee
Therefore \be\label{thetaeq} \theta(\rho;p)= 2|k|p\int\de s\,\frac{1}{\sinh^2\rho(s)}=\pm
2p\int\de\rho\,\frac{1}{\sinh\rho\sqrt{\sinh^4\rho-p^2}}.\ee
The limit $\rho\to\infty$ yields the boundary values of $\theta$, that is the fiber positions on the base $S^2$: if on the negative branch we set $\theta(\infty;p)=\theta_1(p)$, while
$\theta(\infty;p)=\theta_2(p)$ on the positive one, we have\be
(\theta_2-\theta_1)(p)\equiv\Delta\theta(p)=4p\int_{\rho_0(p)}^\infty\de\rho\,\frac{1}{\sinh\rho\sqrt{\sinh^4\rho-p^2}},\ee
which is indeed a (non-trivial) relation between the
angular separation of the boundary fibers and the parameter $p$. The
integral above can be evaluated explicitly and we get (with the same notations as before)
\be
\label{dtheta}
\begin{split}
\Delta\theta(p)=&4p\int_{\sqrt{1+p}}^\infty\de
x\,\frac{1}{(x^2-1)\sqrt{(x^2-1)^2-p^2}}\\
=&4p\int_{\sqrt{1+p}}^\infty\de
x\,\frac{1}{(x^2-1)\sqrt{(x^2-b^2)(x^2\pm a^2)}}=4pI_2(p),
\end{split}
\ee 
where 
\be
I_2(p)\equiv\left\{\begin{array}{ll}\frac{1}{b}\left(\Pi(\frac{1}{b^2}|\frac{a^2}{b^2})-K(\frac{a^2}{b^2})\right)&\textrm{ for } a^2=1-p>0\\
&\\
\frac{1}{(a^2+1)\sqrt{a^2+b^2}}\left(\Pi(\frac{a^2+1}{a^2+b^2}|\frac{a^2}{a^2+b^2})-K(\frac{a^2}{a^2+b^2})\right)
&\textrm{ for } a^2=p-1>0.\end{array}\right.\ee
\begin{figure}[!h]
\centering\includegraphics[width=8cm]{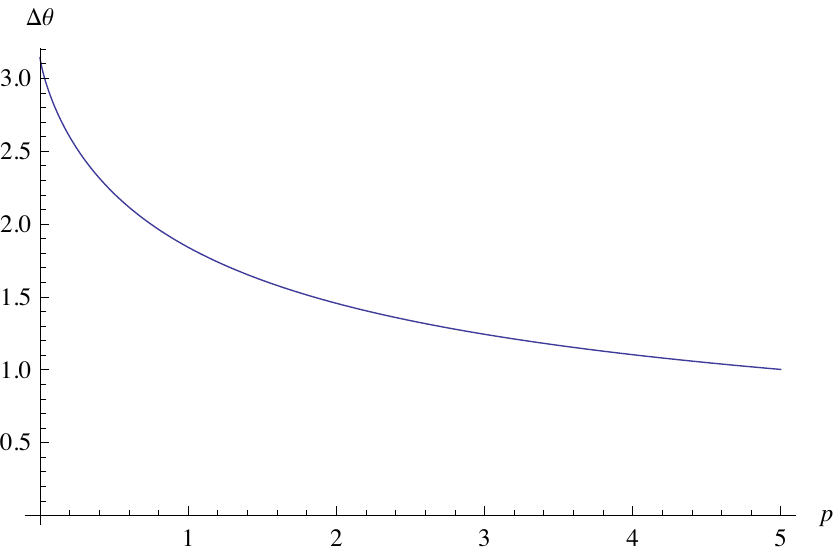} \caption{\footnotesize Plot of $\Delta
\theta(p)$. $\Delta\theta(0)=\pi$, $\Delta\theta(\infty)\to 0$.}\label{figDTheta}
\end{figure}From the plot in Fig.\ref{figDTheta}, we can see that at the maximal
meaningful separation ($\Delta\theta=\pi$) we have a disconnected
solution ($p=0$). The other limit $p\to\infty$ corresponds to
coincident fibers. In a sense, there is a sort of
continuous\footnote{We mean that the disconnected solution can be
continuously obtained from the connected one by going continuously
from $p>0$ to $p=0$. As we will explain, this corresponds to a
BPS/non-BPS transition.} ``phase transition'' at $p=0$. A similar
phenomenon occurs in the correlator of two parallel and concentric
circles (Gross-Ooguri phase transition). However, in that case it
exists a region where the connected solution has area bigger than
the disconnected one, and beyond a certain separation it ceases even
to exist. In our case this situation never happens.

From \reff{rhoequations} it follows that the implicit form of $\rho(s)$ is given by
\be\label{seq} (s-s_0)(p)=\frac{1}{|k|}\int_{\sqrt{1+p}}^{\cosh\rho}\de x \frac{1}{\sqrt{(x^2-b^2)(x^2\pm a^2)}},\ee
where $s_0(p)$ represents the turning point $\rho'(s_0)=0$, $\rho(s_0)=\rho_0$. The range of the parameter $s$ can be determined by taking the limit $\rho\to+\infty$
\be
\begin{split}
\Delta s(p)\equiv (s_2-s_1)(p)=&\frac{2}{|k|}\int_{\sqrt{1+p}}^{\infty}\de x \frac{1}{\sqrt{(x^2-b^2)(x^2\pm a^2)}}\\
=& \frac{2}{|k|}\left\{\begin{array}{ll}\frac{1}{b}K(\frac{a^2}{b^2})&\text{ for } a^2=1-p>0\\
\frac{1}{\sqrt{a^2+b^2}}K(\frac{a^2}{a^2+b^2})&\textrm{ for } a^2=p-1>0.\end{array}\right.
\end{split}
\ee
\begin{figure}
{\includegraphics[width=7cm]{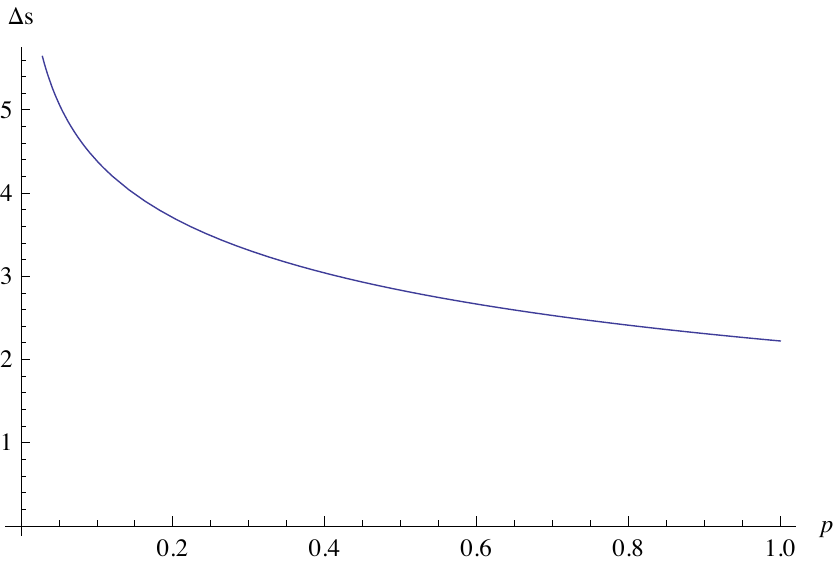}}
{\includegraphics[width=7cm]{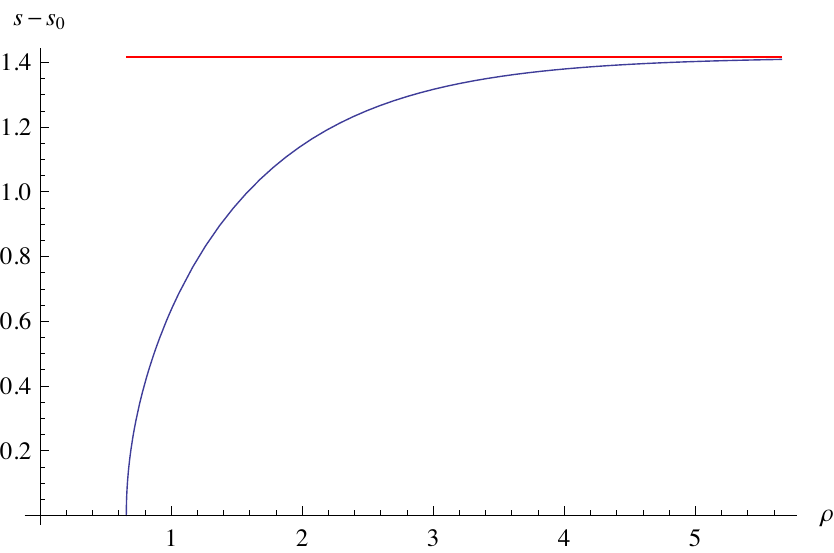}} \caption{\footnotesize Plots
of $\Delta s(p)$ and $(s-s_0)(\rho)$ at $p=0.5$. According to
\reff{rho12}, $\Delta s$ diverges at $p=0$. The red line represents
the boundary value of $s$.}\label{rhoeq}
\end{figure}This is shown in Fig.\ref{rhoeq}. For $p\to 0$ the range diverges, corresponding to the fact that the turning point is mapped to the deep interior $\rho\to 0$ of $AdS_5$. Actually, $p=0$ describes the disconnected solution.

\subsection{Motion in $S^5$}
\begin{figure}
{\includegraphics[width=7cm]{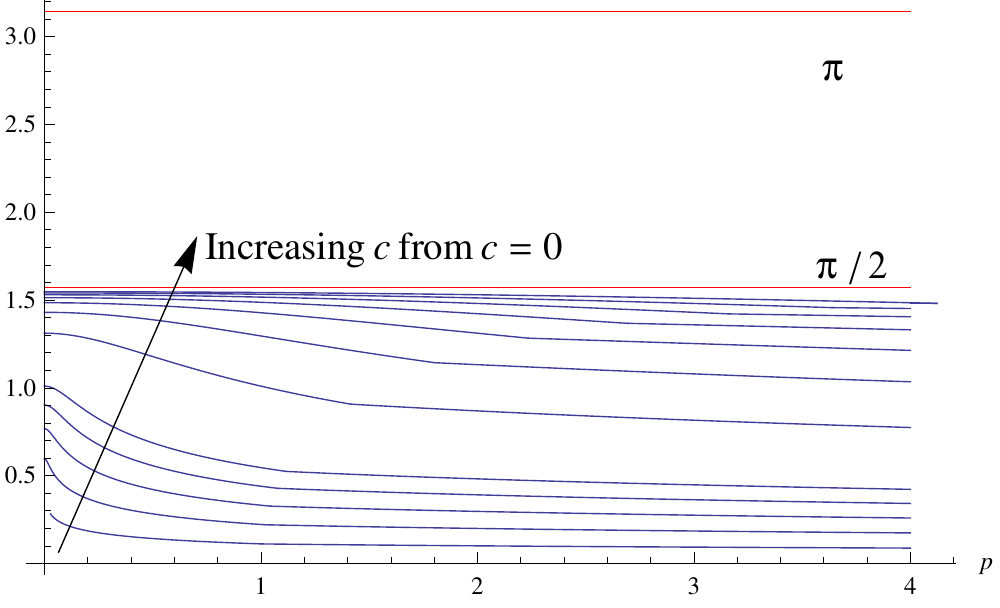}}
{\includegraphics[width=7cm]{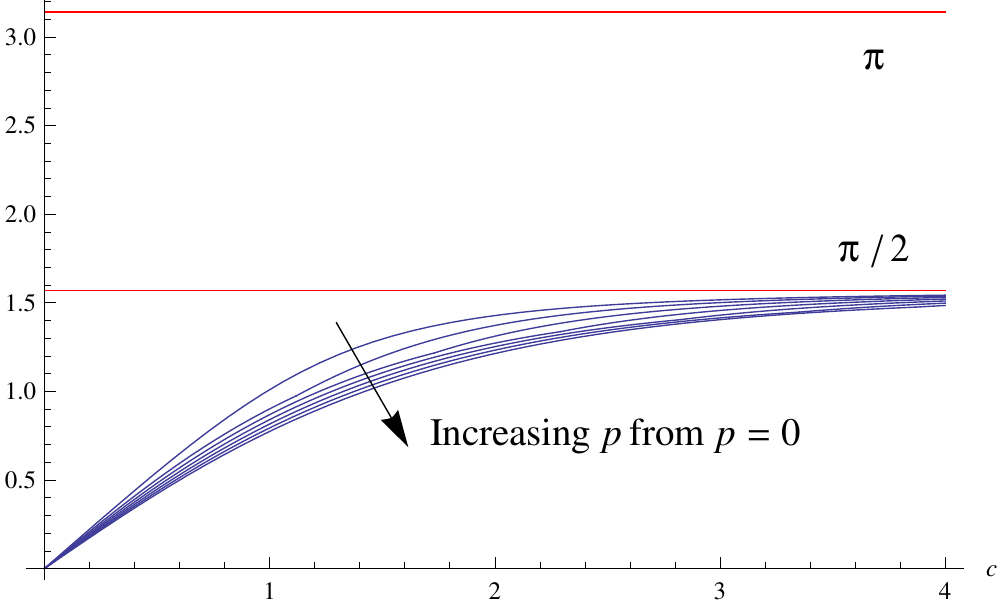}}
\caption{\footnotesize Constant $p$, $c$ lines of $cI_3(p,c)$.}\label{figPC}
\end{figure}

We will now consider what happens when we allow an $S^5$ motion, namely when $c^2>0$.
From \reff{rhoequations} we can see that the area \reff{area2} is simply
replaced by \be
\frac{A_2(c,p)}{4|k|\pi}=\int_{\rho_0(c,p)}^\infty\de\rho\,\frac{\sinh^3\rho}{\sqrt{\sinh^4\rho-c^2\sinh^2\rho-p^2}},\ee
where $\rho_0(c,p)$ is the unique real positive zero of the square
root as determined in \reff{turningpoints}. By
setting\be x=\cosh\rho\qq p'=\sinh^2\rho_0(c,p)\qq B=\sqrt{1+p'}\qq
A=\sqrt{\pm(p'-1-c^2)},\ee the area is given as in
\reff{area2int}, \reff{area2ell} but with the replacements
$b\rightarrow B$, $a\rightarrow A$. In the following we must be careful
about the range of the parameter $s$: in fact, it is determined by
the $\rho$ equation, but it is also the parameter of the $S^5$ motion. From \reff{rhoequations} we get
\be
(s_2-s_1)(p,c)\equiv\Delta s(p,c)=\frac{2}{|k|}\int_B^\infty\de
x\,\frac{1}{\sqrt{(x^2-B^2)(x^2\pm A^2)}}=\frac{2}{|k|}I_3(p,c),\ee
where $\rho(s\to s_{1,2})\to\infty$ and \be
I_3(p,c)\equiv\left\{\begin{array}{ll}\frac{1}{B}K(\frac{A^2}{B^2})&\textrm{
for
}A^2=1+c^2-p'>0\Rightarrow p^2-c^2<1\\
&\\
\frac{1}{\sqrt{A^2+B^2}}K(\frac{A^2}{A^2+B^2})&\textrm{ for }
A^2=p'-1-c^2>0\Rightarrow p^2-c^2>1.\end{array}\right.\ee The
parameters $p$ and $c$ are not independent of each other. Rather,
they must satisfy a compatibility condition. In fact, when the
fibers have the same (opposite) scalar charge \reff{scalarcharge}
they sit on the same (antipodal) point(s) of $S^5$ at the AdS
boundary. Since we have considered an $S^5$ motion given by
\reff{s5motion} \be \beta(s)=|k|c(s-s_0)+\beta_0,\ee we must have $\beta(s_2)=\beta(s_1)\mod 2\pi$ ($\beta(s_2)=\beta(s_1)+\pi\mod 2\pi$), where
$\beta(s_{1,2})$ represent the boundary values of the $S^5$ angle.
Then the compatibility condition \be\label{compatibility} |k|c\Delta
s(p,c)=n\pi\Rightarrow c I_3(p,c)=n\frac{\pi}{2}\qq n\in
2\mathds{Z}\text{ }(2\mathds{Z}+1)\ee must be satisfied. We remark that it is a very
strong constraint on the full $AdS_5\times S^5$ motion, and it turns
out there is no solution other than $n=c=0$, as shown in
Fig.\ref{figPC}. As a matter of fact, the function on the left-hand side of
\reff{compatibility} reaches the minimum required value only
asymptotically for $c\to\infty$. We conclude that there exists no connected
solution for two antiparallel Hopf fibers of opposite scalar charge
(at least as far as the made ansatz is concerned).
\\

So far, the study of the Hopf fiber correlator has been (almost) purely geometrical, and this is motivated by the AdS/CFT geometric prescription to calculate Wilson loop expectation values. However, the string dual of a supersymmetric system of DGRT loops is expected to be represented by a supersymmetric string surface, and then the surface has to be a solution of the sigma model as well as of the $\kappa$-symmetry equations. In the following section we are going to explore this issue for the connected solution, and we will show that it is not supersymmetric.

\section{Supersymmetry}
\label{secSUSY}
In order to verify whether the found solution is a supersymmetric string surface dual to a correlator of two DGRT Hopf fibers, we follow an argument proposed in \cite{dgrta} (see also \cite{Dymarsky:2009si} for a general discussion). We will be rather sketchy, and we will refer to that paper and references therein for details.

In $AdS_5\times S^5$ the $\kappa$--symmetry equations read as
\be\label{ksymm} \left(\ep^{ab}\pa_a X^M \pa_b X^N \Gamma_{MN}- i G_{MN}\pa_a X^M \pa_b X^N \gamma^{ab}\sqrt{\gamma}\right)\ep_\text{AdS}(X)=0,\ee
where $\Gamma^{M}$ are the 10-dimensional curved space gamma matrices, $\ep^{ab}$ is the 2-dimensional antisymmetric tensor and $\ep_\text{AdS}(X)$ is the $AdS_5\times S^5$ conformal Killing spinor.

To check the supersymmetry of the connected solution, we should plug it into \reff{ksymm}. However, there is a simpler way to perform such a verification. Let us consider the following equations
\be
\label{holoeq}
\begin{split}
&V_a^M\equiv \pa_a X^M-\kappa J^M_{\phantom{M}N}j_a^{\phantom{a}b}\pa_b X^N=0\qq X^M\in AdS_4\times S^2\subset AdS_5\times S^5,\\
&\kappa=\pm 1\qq j_a^{\phantom{a}b}=\frac{1}{\sqrt{\gamma}}\gamma_{ac}\ep^{cb}\qq \ep^{12}=1=-\ep_{12},
\end{split}
\ee
where $j_a^{\phantom{a}b}$ is the canonical complex structure associated to the world-sheet metric and $J^M_{\phantom{M}N}$ is a certain matrix defined in the subspace $AdS_4\times S^2$ where the solution lives\footnote{We remind that $\tau=0$.}. It defines an almost complex structure ($J^2=-\mathds{1}$), and represents an elegant way to collect the BPS properties of DGRT loops. A surface extending into $AdS_4\times S^2$ and satisfying $V^M_a=0$ defines a pseudo-holomorphic surface with respect to $J$. Since $J$ is related to the BPS nature of DGRT loops, dual string surfaces are expected to be compatible with it, i.e. pseudo-holomorphic with respect to $J$. In support of this conjecture, in \cite{dgrta} it has been shown that classical pseudo-holomorphic string surfaces are automatically supersymmetric: we are interested in verifying whether the found string solution is pseudo-holomorphic. Taking the explicit form of $J$ as given in \cite{dgrta} (rewritten in terms of the coordinates we have employed), we can plug the given ansatz into the pseudo-holomorphic equations \reff{holoeq}, and the resulting independent conditions turn out to be
\be \rho'+\kappa k\sinh\rho=0\qq \theta'=\phi'=\eta'=0\qq \eta'+\phi'\cos\theta=0.\ee
We can recognize the equations associated to the 1/2 BPS circular loop ($p=0$) and the second Virasoro constraint, which is however automatically satisfied by the requirement $\theta'=\phi'=\eta'=0$. Thus the pseudo-holomorphic equations single out the disconnected solution\footnote{We have verified that the same result can be directly obtained from the $\kappa$--symmetry equations.}.

We observe that the connected string solution we found is not describing the connected correlator of two DGRT Hopf fibers: the non-supersymmetric nature of the solution can be traced back to a ``wrong'' relative orientation of the boundary circles. To see this, let us come back to the definition of a DGRT loop \reff{dgrtscalar}
\be W(\gamma,\theta)=\tr \pexp\left( i\oint_\gamma \de t\;\Big( \dot x^\mu A_\mu+i\dot x^\mu x^\nu\sigma_{\mu\nu}^i\phi_i\Big)\right).\ee When $\gamma$ is a Hopf fiber \reff{hopfcoord} parameterized by $\psi(t)=2kt$, the above operator becomes
\be W(\psi,k)=\tr \pexp\left( i\int_{0}^{2\pi} \de t\;\Big( \dot x^\mu A_\mu-ik\phi_3\Big)\right)\ee because (with the conventions of the previous sections)
\be \theta,\phi=const\qq\psi(t)=2kt \Rightarrow \theta^I=\text{sign}(k)\delta^I_3.\ee
In considering the correlator $\langle W(\psi_1,k_1)W(\psi_2,k_2)\rangle$ of two Hopf fibers, the BPS configurations involve equally oriented fibers with the same scalar charge, or oppositely oriented fibers of opposite scalar charge. This observation suggests that our non-BPS connected string solution describes the correlator of two oppositely oriented Hopf fibers with the same scalar charge.

In fact, the non-supersymmetric connected solution is characterized by having no motion in $S^5$, implying that the fibers have the same scalar charge. Moreover, the tangential direction $\pa_t$ and the radial direction\footnote{$\rho=\rho(s)$.} $\pa_s$ along one fiber on the AdS boundary define an orientation that must be preserved along the world-sheet. However, in the two different radial branches the sign of $\rho'$ is opposite, and hence also the sign of the tangential vector has to be reversed to keep the orientation. We get the picture of a pair of antiparallel Hopf fibers with the same scalar charge.

Interestingly enough, our solution is able to interpolate between a BPS configuration and a non-BPS one. We saw that for general values of $p$ (or $\Delta\theta$) the string surface is not supersymmetric, but for $p\to 0$ (or $\Delta\theta\to\pi$) the solution approaches the disconnected one made of a pair of surfaces which are separately $1/2$ BPS (single great circle). When considered together, one surface halves the preserved supersymmetries of the other, giving rise to a $1/4$ BPS configuration. This situation can be easily understood from the gauge theory viewpoint: the fibers at $\theta=0$ and $\theta=\pi$ lay on the plane $(x^3,x^4)$ and $(x^1,x^2)$ respectively. Each loop represents the $1/2$ BPS great circle, and the presence of a second circle in the orthogonal plane simply halves the components of the independent supersymmetric spinor by imposing a chirality condition. The explicit calculation can be found in \cite{dgrta} for the case of two DGRT parallel fibers, but the same applies also to the non-DGRT antiparallel fibers in the special case $\theta_1=0$, $\theta_2=\pi$, since the supersymmetry variation becomes insensible \footnote{The only difference is a flip of the chirality.} to the ``wrong'' sign of the charges.

\section{ Quark-antiquark static potential}\label{secpot}
The connected solution we have found represents a system made of a pair of antiparallel fibers, with the same scalar charge: in some sense, it can be thought of as a topologically non-trivial analogue in $S^3$ of the antiparallel lines describing the static quark-antiquark potential. In fact, when the fibers are sufficiently close to each other (namely $\Delta\theta\to 0$, or $p\to\infty$), locally they look like straight antiparallel lines. In this picture, we can expect that the role of the separation distance $L$ between the lines in the (strong coupling) static potential \cite{pot1,mwloop}
\be V(L)=-\frac{4\pi^2}{\Gamma^4(1/4)}\frac{\sqrt{\lambda}}{L}\ee
will be played by $\Delta\theta/2$ (in the large $p$ limit), the distance between the Hopf fibers on the base in the $S^3$ metric. This will indeed be the case. 


There is a subtle point in this interpretation: while the two antiparallel lines are thought as a degenerate rectangular single loop of infinite extension, involving only one trace, we are dealing here with the correlator of two loops, involving two different traces. The precise relation between the two quantities is well known, involving the physical interpretation of the different loop correlators in terms of singlet and adjoint potentials \cite{Brown:1979ya,Nadkarni}.

A quark-antiquark pair at the same point can be in either a singlet or an adjoint state, according to the irreducible representations of the tensor product
\be\label{nnbar}
\bar{N}\otimes N = 1 \oplus ( N^ 2 -1 ). 
\ee
The usual strategy to extract the quark-antiquark potential from Wilson loop is to relate the four point function (in the infinite mass limit)
\be
G(x_1,x_2;y_1,y_2)=\langle 0|T(\bar{Q}(x_1)Q(x_2)Q^{\dagger}(y_2)\bar{Q}^{\dagger}(y_1))|0\rangle
\ee
to the gauge invariant phases, experienced by the fields $Q$'s in their temporal evolution, as $T\to\infty$. Here $x_{1,2}$ are placed at $T/2$ while $y_{1,2}$ at $-T/2$ (see \cite{Brown:1979ya,Nadkarni} for further details). The field $Q(x)$ has a color index and the general structure of the correlation function in the large $T$ limit, taking $\vec{x}_1=\vec{y}_1$ and $\vec{x}_2=\vec{y}_2$, is
\be
G(\vec{x}_1,\vec{x}_2;T)=\mathbb{P}_S \exp[-TV_S(\vec{x}_1-\vec{x}_2)] +\mathbb{P}_A\exp[-TV_A(\vec{x}_1-\vec{x}_2)].
\ee
The matrix operators $\mathbb{P}_S$ and $\mathbb{P}_A$ project respectively onto the singlet state and the adjoint state, according to the decomposition (\ref{nnbar}). By taking the relevant traces with the projectors and relating the four-point functions to correlators of Wilson lines we get the following identities:
\be
 \exp[-TV_S(\vec{x}_1-\vec{x}_2)]=\frac{1}{N}\langle{\rm Tr}[W(\vec{x}_1)W^\dagger (\vec{x}_2)]\rangle
 \ee
and
\be
\exp[-TV_A(\vec{x}_1-\vec{x}_2)]=\frac{1}{N^2-1}\langle{\rm Tr}[W(\vec{x}_1)]{\rm Tr}[W^\dagger (\vec{x}_2)]\rangle-\frac{1}{N(N^2-1)} \langle{\rm Tr}[W(\vec{x}_1)W^\dagger (\vec{x}_2)]\rangle,\ee
$W(\vec{x})$ being the Wilson line in $\vec{x}$ extending along the Euclidean time. The first relation is the familiar definition of the quark-antiquark potential $V_S$ in terms of antiparallel Wilson lines. The second equality gives us instead a physical interpretation of the correlator two traced Wilson lines, in terms of the singlet and adjoint potential $V_A$. Using the explicit definition of the singlet potential we finally arrive at
\be \frac{N^2-1}{N^2}\,e^{-TV_A }=W(\vec{x}_1,\vec{x}_2)-\frac{1}{N^2}\,e^{-TV_S }\ee
that expresses the normalized connected correlator $W(\vec{x}_1,\vec{x}_2)$ of the two traced Wilson lines in terms of the potentials. This is the key relation to derive the quark-antiquark potential, in the limit of small separation between the fibers, from the DGRT loops correlator both at weak and strong coupling.

We proceed as follows: assuming for $W$ an expansion in powers of $T$, we have up to the second order in $T$
\begin{align} W_0&=1, \nn\\ V_A&=-\frac{1}{N^2-1}\left(V_S + N^2 W_1\right),\label{VaPot} \\
V_S&=-W_1\pm\sqrt{(N^2-1)(2 W_2-W_1^2)}.\label{VfPot}\end{align}
In order to keep track of the powers of $N^2$, we set $\hat{V_A}=N^2 V_A$, $\hat{W_1}=N^2 W_1$ and $\hat{W_2}=N^2 W_2$, so that $\hat{V_A}\sim\hat{W_1}\sim V_S$ and $\hat{W_2}\sim V_S^2$ and the  \eqref{VfPot} becomes
\be V_S^2+\frac{2\hat{W_1}}{N^2}V_S+\frac{\hat{W_1}^2}{N^2}-\frac{2(N^2-1)}{N^2} \hat{W_2}=0.\ee
The solution to this equation can be written as a power series of $\frac{1}{N^2}$
\be V_S=-\sqrt{2 \hat{W_2}}+\frac{1}{N^2}\left(\frac{\sqrt{2\hat{W_2}}}{2}+\frac{\hat{W_1}^2}{\sqrt{2\hat{W_2}}}-\hat{W_1}\right)+O(N^{-4}),\ee
and, in the large $N$ limit, we can expand in power of $g^2$ finding 
\be V_S=-\sqrt{2 W_2^{(2)}}\,g^2-\frac{W_2^{(3)}}{\sqrt{2W_2^{(2)}}} g^4.\ee
This simple result shows that if we want to compute the $g^4$ order of the static potential, we need the $g^6$ order for the correlator of Wilson loops. However, the technical difficulties in computing Feynman graphs allows only for numerical computations, as in the case of \cite{twotwo}. We will not attempt this computation and we will only check the leading order result.

At strong-coupling, due to the exponentiation property of the Wilson loops correlator, we can directly relate our string solution with the singlet potential: let us compute the relevant quantities in the large $p$ limit. The regularized area \reff{area2reg} becomes
\be A_2^R=\Delta A(p)= A_0\sqrt{p}\qq A_0\equiv 4\pi\left(\frac{1}{\sqrt{2}}K(1/2)-\sqrt{2}E(1/2)\right)=-\frac{4\sqrt{2}\pi^{5/2}}{\Gamma^2(1/4)}\ee
and in the same limit the angular separation \reff{dtheta} approaches
\be\Delta\theta(p)=\frac{\theta_0}{\sqrt{p}}\qq\theta_0\equiv 2\sqrt{2}\left(\Pi(1/2,1/2)-K(1/2)\right)=\frac{4 \sqrt{2} \pi ^{3/2}}{\Gamma^2(1/4)},\ee
obtaining
\be \sqrt{\lambda}S^R_\text{min}(p)=\frac{\sqrt{\lambda}}{2\pi}\Delta A(p)=-\frac{16 \pi^3}{\Gamma^4(1/4)}.\ee
Since
\be \langle WW(A^R_2)\rangle\simeq e^{-\frac{\sqrt{\lambda}}{2\pi}\Delta A(p)},\ee
we naturally define
\be \wt V\equiv -\frac{1}{T}\ln\langle WW(A^R_2)\rangle\qq \Delta\theta/T\ll 1,\ee
where $T=2\pi$ is the length of the fiber in the $S^3$ metric. We see immediately that
\be \wt V= -\frac{4\pi^2}{\Gamma^4(1/4)}\frac{\sqrt{\lambda}}{\Delta\theta/2}\ee
which, as expected, coincides with the function $V(L)$ once we identify $L\leftrightarrow \Delta\theta/2$.

The same result is manifest also at weak coupling, where at order $\lambda$ we have (appendix \ref{apppert})
\be \langle W(\psi_\uparrow)W(\psi_\downarrow)\rangle_\text{conn}=
\frac{\lambda}{2}\left(\frac{1}{\sin\left(\Delta\theta/2\right)}-1\right)\simeq
\frac{\lambda}{4\pi}\frac{2\pi}{\Delta\theta/2}\ee
in agreement with the perturbative calculation for the antiparallel lines \cite{pot2,twotwo}
\be \langle W(\uparrow)W(\downarrow)\rangle_\text{conn}=\frac{\lambda}{4\pi}\frac{T}{L}
\qq \frac{T}{L}\to\infty\ee
provided we identify $T=2\pi$ and $L\leftrightarrow\Delta\theta/2$, as before.    

\section{Conclusions}\label{secconcl}
In this paper we have studied at string level the correlator of two Wilson loops involving Hopf fibers of $S^3$ in $\N=4$ SYM theory. The system has been originally introduced in the context of Wilson loops in \cite{dgrta}, and it represents an example of topologically non-trivial loop configuration. Our main motivation for the study of such a system was indeed to explore non-trivial topologies, and to continue the analysis of configurations involving circles. The study has been performed at strong-coupling, using AdS/CFT correspondence, and we have been able to find a connected string surface linking the two boundary fibers. It has been shown that the solution describes the connected correlator of a pair of antiparallel Hopf fibers with the same scalar charge. For this reason, we have observed that such a system can be thought of as a deformation of the ordinary antiparallel lines describing the static quark-antiquark potential. Indeed, we have verified at weak and strong coupling that the latter system is recovered in the small separation limit, once the fiber distance is identified with distance between the lines. We observed an interesting dynamics, where the string solution is able to continuously interpolate between a non-BPS configuration and a BPS one by varying a parameter.

Moreover, our analysis suggests that the correlator of two antiparallel DGRT Hopf fibers does not admit a connected string solution, and presumably the parallel ones neither. Of course, this is certainly true as far as the considered type of solution is concerned, but it provides by no means a general proof. We can only observe that the our ansatz is the most natural and minimal one with respect to the Hopf fibration geometry: it seems very difficult to construct an inevitably more complicated solution respecting the desired boundary conditions. This fact may be considered as a string theory result in favor of the exactness of the matrix model for two DGRT fibers. It would be interesting, of course, to check this expectation directly in the gauge theory side at higher order in perturbation theory.

Another interesting direction worth to be studied is the application of the found string solution to the holographic calculation of the 3-point correlation function of semiclassical states and non-BPS operators \cite{jsw,zarembo,Costa:2010rz}. Even though a precise string description of such observables is still a long way off, recently the problem has been faced in the simplified version of considering the correlator of two heavy operators and a light one. In some cases two Wilson loops have been considered as the heavy operators \cite{aldaytseytlin}, and then the semiclassical computation of the 3-point function amounts to evaluate the string vertex operator of the light operator over the classical world-sheet generated by the heavy ones. For such calculations an explicit and analytical expression of the classical string solution is needed, but only few are known. In this paper we have presented a new type of solution, and thus it can be used in that context. We defer this possibility to future investigations. 
\section*{Acknowledgements}
This work was supported in part by the MIUR-PRIN contract 2009-KHZKRX.  We thank Nadav Drukker for illuminating discussions and Donovan Young for sharing with us his perturbative results.

\newpage
\appendix
 \begin{flushleft}
 {\huge \bf Appendices}\hfill
 \end{flushleft}
\addcontentsline{toc}{section}{\large Appendices}
\renewcommand{\theequation}{\Alph{section}.\arabic{equation}}
\section{Explicit form of the solution}\label{appint}
For completeness, we report here the explicit form of the functions
$\rho(s)$, $\theta(s)$, $\phi(s)$, $\eta(s)$. From \reff{seq} we
have
\be\pm|k|(s-s_0)=\left\{\begin{array}{ll}\frac{1}{\sqrt{1+p}}F\left(\mu|\frac{1-p}{1+p}\right)&\text{ for } 1-p>0\\
\frac{1}{\sqrt{2p}}F\left(\ep|\frac{p-1}{2p}\right)&\textrm{ for } p-1>0,\end{array}\right.\ee
where
\be \mu=\arcsin\sqrt{\frac{\cosh^2\rho-1-p}{\cosh^2\rho-1+p}}\qq \ep=\arccos\frac{\sqrt{1+p}}{\cosh\rho}.\ee
Therefore\footnote{$y=F(\phi|m)$, $\text{sn}(y)=\sin(\phi)$, $\text{cn}(y)=\cos(\phi)$.}
\be\cosh\rho(s)=\left\{\begin{array}{ll}
\sqrt{\frac{(p-1)\text{sn}^2\left(|k|\sqrt{1+p}(s-s_0)|\frac{1-p}{1+p}\right)+1+p}{1-\text{sn}^2\left( |k|\sqrt{1+p}(s-s_0)|\frac{1-p}{1+p}\right)}}&\text{ for } 1-p>0\\
\frac{\sqrt{1+p}}{\text{cn}\left(|k|\sqrt{2p}(s-s_0)|\frac{p-1}{2p}\right)}&\text{
for } p-1>0.\end{array}\right.\ee We see that it is much more
convenient to express all the variables as functions
of\footnote{Reminding the relation between global and Poincar\'e
coordinates $(z,x^\mu$) of $AdS_5$ \[ z=\frac{e^\tau}{\cosh\rho}\qq
x^\mu=\tanh\rho e^\tau \Omega^\mu\qq \Omega^2=1\qq \de s^2=\frac{\de
z^2+\de x^\mu\de x^\mu}{z^2}\] one can also easily trade $z$ for
$\cosh\rho$.} $\rho$. As far as the angular variables $\theta$ are
concerned, we start from the case $\phi'=\eta'=0$. From
\reff{thetaeq} we get \be\label{thetaeqapp} \theta-\theta_0=\pm
2p\int_{\sqrt{1+p}}^\rho\de\rho\,\frac{1}{\sinh\rho\sqrt{\sinh^4\rho-p^2}}=\pm2pI(\rho,p),\ee
where \be I(\rho,p)=\left\{
\begin{array}{ll}\frac{1}{p\sqrt{p+1}}\left(2\Pi(-1,\mu|\frac{1-p}{1+p})-F(\mu|\frac{1-p}{1+p})\right)&\text{ for } 1-p>0\\
&\\
\frac{1}{p\sqrt{2p}}\left((1+p)\Pi(-\frac{1}{p},\ep|\frac{p-1}{2p})-pF(\ep|\frac{p-1}{2p})\right)&\text{ for }p-1>0\end{array}\right.\ee
and the value $\theta_0$ at the turning point corresponds to the mean value $(\theta_1+\theta_2)/2$.

As observed in Section \ref{secstringhopf}, in principle it is
possible to work with $\phi',\eta'\neq0$. In this case, the relevant
equations are \reff{phietaeq}, \reff{thetaeqbis} (we take
$c_\phi>0$) \be\label{thetaeqapp02} \eta'=-\phi'\cos\theta\qq
\phi'=\frac{|k|c_\phi}{\sin^2\theta\sinh^2\rho}\qq\theta'^2=\frac{4k^2p^2}{\sin^2\theta\sinh^4\rho}\left(\sin^2\theta-\frac{c_\phi^2}{4p^2}\right)\ee
and \reff{rhoequations}
\be\rho'=\pm\frac{|k|}{\sinh\rho}\sqrt{\sinh^4\rho-p^2}.\ee
Consistency requires $\sin\theta_0\equiv\frac{c_\phi}{2p}\leq 1$,
$\theta_0\leq\theta\leq\pi-\theta_0$, and the extremum ($\theta'=0$)
is taken in correspondence of $\rho_0$
($\theta'=\rho'\pa_\rho\theta$). 
Assuming it is located at $\theta_0<\pi/2$, from \reff{thetaeqapp02}
it follows 
\be\label{thetaeqapp01}
\int_{\frac{\cos\theta}{\cos\theta_0}}^{1}\frac{\de
Y}{\sqrt{1-Y^2}}=\arccos\left(\frac{\cos\theta}{\cos\theta_0}\right)=\pm2p
I(\rho,p).
\ee 
In the case $c_\phi=0$, the expression
\reff{thetaeqapp} is recovered (except for the integration constant,
which does not represent any extremum for $\theta$). For
$\theta_0=\pi/2$, the left-hand side of \reff{thetaeqapp01} is not
well defined. However, from \reff{thetaeqapp02} we can realize that
it is consistent to consider $\theta=\theta_0$, $c_\phi=2p$, and
then $\theta$, $\phi$ simply exchange the roles with respect to the
case $c_\phi=0$.

Now that $\theta$ is known, the $\phi$, $\eta$ equations can be
easily integrated 
\be
\begin{split}
\frac{\de\phi}{\de\rho}=&\pm\frac{c_\phi}{\sin^2\theta}\frac{1}{\sinh\rho\sqrt{\sinh^4\rho-p^2}}=\frac{1}{\sin\theta\sqrt{\sin^2\theta-\sin^2\theta_0}}\frac{\de\theta}{\de\rho}\\
\phi-\phi_0=&\frac{\pi}{2}-\arctan\left(\frac{\sqrt{2}\cos\theta\sin\theta_0}{\sqrt{\cos(2\theta_0)-\cos(2\theta)}}\right)\\
\frac{\de\eta}{\de\rho}=&-\frac{\cos\theta}{\sin\theta\sqrt{\sin^2\theta-\sin^2\theta_0}}\frac{\de\theta}{\de\rho}\\
\eta-\eta_0=&-\arctan\left(\frac{1}{\sin\theta_0}\sqrt{\frac{\cos(2\theta_0)-\cos(2\theta)}{2}}\right)
\end{split}
\ee
We conclude by observing that there is another derivative that can
be turned on, namely $\tau'$. Throughout the paper we have
considered $\tau'=0$, and this is motivated by the fact that we
wanted to consider fibers on the same 3-sphere. Therefore, on the
AdS boundary, $\tau$ has to be the same for the two fibers, and the
simpler way to guarantee this is indeed to set $\tau'=0$. 

\section{Perturbative calculations}\label{apppert}
We present here the perturbative expansion, up to order $\lambda^2$, of the connected correlator $\langle W(\psi_1)W(\psi_2)\rangle_\text{conn}$ of two antiparallel Hopf fibers ($\psi_1=2t,\psi_2=-2s$) of equal scalar charge ($\theta_1^I\theta_{2I}=1$). The parameters along the two fibers will be denoted by $t$ and $s$, $t,s\in[0,2\pi]$, while points will be denoted by $x$ and $y$. Reminding the general expression of the propagator \reff{circprop}, evaluating it on different fibers\footnote{Due to the relation between the orientation and the scalar charge, the propagator is no longer constant.} we get
\be G^{ab}(x,y)=\frac{g^2\delta^{ab}}{8\pi^2}\left[\frac{2}{1-\cos (u) \cos \left(\frac{\Delta \theta }{2}\right) \cos \left(\frac{\Delta \phi }{2}\right)+\sin (u) \sin \left(\frac{\Delta \phi }{2}\right) \cos
   \left(\frac{\theta _1+\theta _2}{2}\right)}-1\right],\ee
where $u=t+s$, $\Delta\theta=\theta_1-\theta_2$, $\Delta\phi=\phi_1-\phi_2$, while
\be G^{ab}(x_1,x_2)=\frac{g^2\delta^{ab}}{8\pi^2},\ee
when the points belong to the same fiber. In the following we are going to consider the particular solution $\Delta\phi=0$, so that the expression above simplifies to
\be G^{ab}(x,y)=\frac{g^2\delta^{ab}}{8\pi^2}\left(f(u)-1\right)\qq f(u)\equiv\frac{2}{1-\cos(u)\cos(\Delta\theta/2)}.\ee
As far as the $U(N)$ gauge group generators $\{T_a\}$ are concerned, we follow the conventions of \cite{twotwo}
\be
\begin{split}
& [T_a,T_b]=if_{abc}T_c\qq a=0,1,\ldots,N^2-1\qq
f_{0bc}=0\\
&\tr (T_a T_b)=\frac{\delta_{ab}}{2}\qq\tr (T_a)=\sqrt{\frac{N}{2}}\delta_{a0}\qq T_0=\frac{\mathds{1}}{\sqrt{2N}}.
\end{split}
\ee

\paragraph{Order $\lambda$.} The only contribution is from the single exchange diagram
\begin{center}
\begin{tabular}{cc}
\mbox{\includegraphics[width=1.6cm,height=1.8cm]{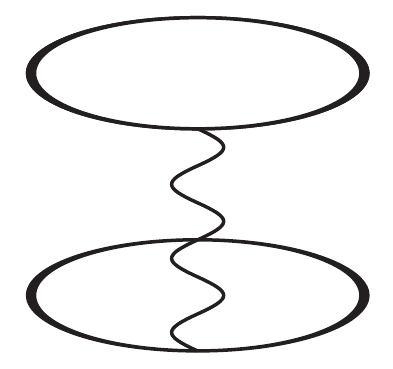}}&\raisebox{.8cm}{$\displaystyle=\frac{g^2}{8\pi^2}\delta^{ab}\tr (T_a)\tr (T_b)\int_0^{2\pi}\de t\de s\,\left(f(u)-1\right)=d^{(2)}$}.\\
\end{tabular}
\end{center}
\noindent Since
\be \delta^{ab}\tr (T_a)\tr (T_b)=\frac{N}{2}\ee
and
\be\label{apppert01} \int_0^{2\pi}\de t \de s\,\left(f(u)-1\right)=(2\pi)^2\left(\frac{2}{\sin(\Delta\theta/2)}-1\right)\equiv F\ee
we get
\be d^{(2)}=\frac{\lambda}{4\pi}(2\pi)\left(\frac{2}{\sin(\Delta\theta/2)}-1\right).\ee

\paragraph{Order $\lambda^2$.} There are two inequivalent diagrams
\begin{center}
\begin{tabular}{cc}
\includegraphics[width=1.6 cm,height=1.8cm]{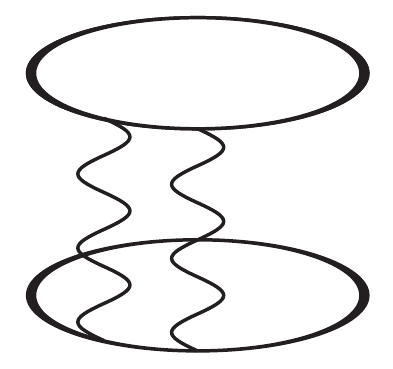}&\raisebox{.8cm}{$\displaystyle=\left(\frac{\lambda}{8\pi^2}\right)^2\delta^{ac}\delta^{bd}\tr (T_a T_b)\tr (T_c T_d)F^2=d^{(4)}_1$,}\\
\includegraphics[width=1.6cm,height=1.8cm]{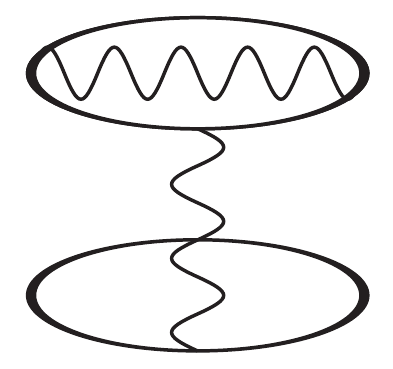}&\raisebox{.8cm}{$\displaystyle\ \ =\left(\frac{\lambda}{8\pi^2}\right)^2 \delta^{ac}\delta^{bd}\tr (T_a T_b T_c)\tr (T_d)(2\pi)^2F=d^{(4)}_2$.}
\end{tabular}
\end{center}

\noindent Since
\be\delta^{ac}\delta^{bd}\tr (T_a T_b)\tr (T_c T_d)=\delta^{ac}\delta^{bd}\tr (T_a T_b T_c)\tr (T_d)=\frac{N^2}{4}\ee
we get
\be d^{(4)}_1=\frac{1}{4}\left(\frac{\lambda}{4\pi}\right)^2(2\pi)^2
\left(\frac{4}{\sin^2(\Delta\theta/2)}-\frac{4}{\sin(\Delta\theta/2)}+1\right)\ee
and
\be d^{(4)}_2=\frac{1}{4}\left(\frac{\lambda}{4\pi}\right)^2 (2\pi)^2\left(\frac{2}{\sin(\Delta\theta/2)}-1\right).\ee

Let us conclude by observing that it is possible to work also with $\Delta\phi\neq0$. In fact, in this case
\be F=(2\pi)^2\left(\frac{4}{d}-1\right)\qq d=\sqrt{2(1-\cos\theta_1\cos\theta_2-\sin\theta_1\sin\theta_2\cos(\Delta\phi)},\ee
where $d$ is the Euclidean distance between two points $(\theta_1,\phi_1)$, $(\theta_2,\phi_2)$ on the base $S^2$. We can see that \reff{apppert01} corresponds indeed to $d|_{\Delta\phi=0}$.

\end{document}